\documentclass[%
 aip,
 amsmath,amssymb,
 reprint,%
]{revtex4-2}

\usepackage{graphicx}% Include figure files
\usepackage{dcolumn}% Align table columns on decimal point
\usepackage{bm}% bold math
\usepackage{color}
\usepackage{xcolor}
\usepackage[utf8]{inputenc}
\usepackage[T1]{fontenc}

\usepackage{caption, subcaption} % Michal Balcerek: for multiple subfigures
\usepackage{multirow}

\begin{document}

\preprint{AIP/123-QED}

% \title[Discrimination procedure for Gaussian processes based on quadratic form test statistics]{Discrimination procedure for Gaussian processes}

% \title[Testing of Gaussian processes with quadratic form statistics]{Testing of Gaussian processes with quadratic form statistics}

%\title[Fractional Brownian motion with random Hurst exponent in a heterogeneous environment]{Fractional Brownian motion with random Hurst exponent in a heterogeneous environment}

\title[Fractional Brownian motion with random Hurst exponent]{Fractional Brownian motion with random Hurst exponent: accelerating diffusion and persistence transitions}

 %\title[Doubly stochastic fractional Brownian motion]{Doubly stochastic fractional Brownian motion in a heterogeneous environment}

% Force line breaks with \\

\author{Michał Balcerek}
\email{michal.balcerek@pwr.edu.pl}
\author{Krzysztof Burnecki}
\affiliation{ 
Faculty of Pure and Applied Mathematics, Hugo Steinhaus Center, Wroclaw University of Science and Technology, Wyspianskiego 27, 50-370 Wroclaw, Poland%\\This line break forced with \textbackslash\textbackslash
}%% \homepage{http://www.Second.institution.edu/~Charlie.Author.}
 \author{Samudrajit Thapa}
  \affiliation{School of Mechanical Engineering, Tel Aviv University, Tel Aviv 6997801, Israel
}%
 \affiliation{Sackler Center for Computational Molecular and Materials Science, Tel Aviv University, Tel Aviv 6997801, Israel}%
\author{Agnieszka Wyłomańska}
\affiliation{ 
Faculty of Pure and Applied Mathematics, Hugo Steinhaus Center, Wroclaw University of Science and Technology, Wyspianskiego 27, 50-370 Wroclaw, Poland%\\This line break forced with \textbackslash\textbackslash
}%% \homepage{http://www.Second.institution.edu/~Charlie.Author.}
\author{Aleksei Chechkin}%
\affiliation{ 
Faculty of Pure and Applied Mathematics, Hugo Steinhaus Center, Wroclaw University of Science and Technology, Wyspianskiego 27, 50-370 Wroclaw, Poland%\\This line break forced with \textbackslash\textbackslash
}%
  \affiliation{Institute for Physics \& Astronomy, University of Potsdam, 14476 Potsdam-Golm, Germany
 %\textbackslash\textbackslash
}%
 \affiliation{Akhiezer Institute for Theoretical Physics, National Science Center ‘Kharkov Institute of Physics and Technology’, Akademicheskaya
st.1, Kharkov 61108, Ukraine}
 
\date{\today}% It is always \today, today,
             %  but any date may be explicitly specified

\begin{abstract}
Fractional Brownian motion, a Gaussian non-Markovian self-similar process with stationary long-correlated increments, has been identified to give rise to the anomalous diffusion behavior in a great variety of physical systems. The correlation and diffusion properties of this random motion are fully characterized by its index of self-similarity, or the Hurst exponent. 
However, recent single particle tracking experiments in biological cells revealed highly complicated anomalous diffusion phenomena that can not be attributed to a class of self-similar random processes. Inspired by these observations, we here study the  process which preserves the properties of  fractional Brownian motion at a single trajectory level, however, the Hurst index randomly changes from trajectory to trajectory. We provide a general mathematical framework for analytical, numerical and statistical analysis of fractional Brownian motion with random Hurst exponent. The explicit formulas for probability density function, mean square displacement and autocovariance function of the increments are presented for three generic distributions of the Hurst exponent, namely two-point, uniform and beta distributions. The important features of the process studied here are accelerating diffusion and persistence transition which we demonstrate analytically and numerically. 

\end{abstract}

\maketitle

\begin{quotation}
Almost 200 years have passed since Robert Brown reported the results of his microscopical observations of very rapid and highly irregular motion of the small pollen particles in the solvent. 
The fundamental origin of this phenomenon, called Brownian motion, was discovered in pioneering works by Einstein, Sutherland and Smoluchowski. Two hallmarks of Brownian motion are the Gaussian  probability distribution of displacements and  linear growth of its variance with time.   
Brownian motion is ubiquitous due to the central limit theorem and naturally appears if evolution of the system or result of experiment is determined by a large number of independent random factors. Mathematically, the (ordinary) Brownian motion  is a self-similar Gaussian process with independent increments. The fractional Brownian motion, introduced by Kolmogorov, is a self-similar Gaussian process whose increments are however strongly correlated.  The correlation and diffusion properties of fractional Brownian motion  are fully characterized by its index of self-similarity, or the Hurst exponent. However,  single particle tracking experiments revealed highly complicated  phenomena in biological cells that can not be explained within the framework of the theory of self-similar processes. {Inspired by several such experiments in viscoelastic environments}, we here study the process which preserves the properties of fractional Brownian motion at a single trajectory level, however, the Hurst exponent randomly changes from trajectory to trajectory. We find that such process exhibits new intriguing properties which we demonstrate analytically and numerically. Our results pave the way for developing consistent mathematical theory of complicated fractional random motions and designing new statistical tools for experimental studies.

\end{quotation}

\section{\label{sec:1}Introduction and motivation}
It has been over a century since Einstein, Sutherland, Smoluchowski and Langevin formulated the theory of Brownian motion
to describe the thermal motion of particles \cite{einstein05,sutherland05,smoluchowski06,langevin08}, {first reported in 1828 by Robert Brown} with his observations of the jiggling motion of pollen grains in water.
The two hallmarks of Brownian motion are the
Gaussian form of the probability density function (PDF) to find a particle at a given position at  fixed time $t$ and the linear growth of its mean squared displacement (MSD), i.e. variance, with time.  The strict mathematical formulation of \textit{ordinary Brownian motion}--- a non-stationary, Gaussian, self-similar process with stationary and independent increments--- highlighted the ubiquity of this random motion in nature due to Central Limit Theorem \cite{kolmogorov49,donsker}. 
% The success of \textit{ordinary} Brownian motion, also known as Wiener process,---a non-stationary, Gaussian, self-similar process with stationary, independent increments--- in describing the motion of diffusive particles 
% has been phenomenal [refs].  
However, many experiments carried out on different scales ranging from astrophysical to inter-cellular ones reported non-linear growth of the $\mathrm{MSD} \propto t^\mu $, $\mu \neq 1$, a phenomenon now known as anomalous diffusion, see e.g. \cite{bouchaud,klages08,inne1,metzler2014} and references therein. 
One of the paradigmatic models of such anomalous diffusion was suggested by Kolmogorov in the context of statistical description of locally homogeneous isotropic turbulence. It comprises a class of Gaussian, self-similar processes possessing stationary power-law correlated increments \cite{kol40}. This model was called \textit{fractional Brownian motion} (FBM) in the seminal paper by Mandelbrot and van Ness \cite{NessMandelbrot} who presented its explicit integral representation and advertised its relevance to a broad scientific community. As motivations for such a generalization, Mandelbrot and van Ness cited examples from economic time series
which exhibited cycles having periods of duration comparable to the sample size (long-range), studies of 
fluctuations in solids (flicker noise), and hydrological experiments where Hurst found ``an infinite interdependence'' between successive water flows (Hurst's law). In the last decades FBM attracted attention in many applied fields, e.g. hydrology, telecommunications, economics, engineering \cite{dok-opp-taqqu}. This is mainly due to its Gaussianity and the power-law behaviour of autocovariance function of its increment process,  which leads to the notion of long-range dependence (long memory) \cite{beran2016long}.

% Many of such experiments could be described by a generalization of the Wiener process to one with power-law correlated increments [Eq. already here? refs]. Such a generalization of the Wiener process to a Gaussian, self-similar process with stationary but power-law correlated increments was already suggested by Kolmogorov in 1940 [ref] and in their seminal papers Mandelbrot and van Ness called it Fractional Brownian motion (FBM) with the power-law correlation of the increments defined by the fixed Hurst exponent $H$ in the range $0<H<1$ [refs]. As motivations for such a generalization, Mandelbrot and van Ness cited examples from economic time-series
% which exhibited cycles having periods of duration comparable to the sample size (long-range), studies of 
% fluctuations in solids (flicker noise), and hydrological experiments where Hurst found "an infinite interdependence" between successive water flows (Hurst's law). 

The anomalous diffusion properties of FBM are characterized by the Hurst exponent $H$, $0<H<1$, which is related to the anomalous diffusion exponent as $H~=~\mu/2$. The {ordinary Brownian motion} is a special case of FBM with $H~=~1/2$. Subdiffusive FBM with $0<H<1/2$ corresponds to anti-persistent motion while superdiffusive FBM with $1/2<H<1$ corresponds to persistent motion thereby making FBM a successful model for diffusion  in complex environments such as a biological cell where, for instance, the viscoelastic environment could lead to anti-persistent motion of organelles while the motion of cellular cargo being actively transported by motor-proteins could be persistent\cite{weiss2012,franosch2013,metzler2014}.  

A number of  modern single particle tracking experiments in heterogeneous and crowded environments---such as those in biological cells---
%\textcolor{red}{combined with  advanced single trajectory analyses methods such as Bayesian inference  and machine learning Samu can we remove this marked text (in red)?} 
have recently reported that while the diffusion of tracked particles is consistent with FBM, {the diffusion exponent and diffusion coefficient extracted from the time averaged mean squared displacement (TAMSD) often vary from trajectory to trajectory.} Examples of such experiments include the dynamics of histonelike nucleoid-structuring proteins\cite{wang2018}, diffusion of membraneless organelles in the single-cell state of C. elegans embryos \cite{benelli2021sub}, diffusion of nanometer sized beads in the biochemically active extracts derived from the eggs of the clawfrog \textit{Xenopus laevis} \cite{speckner2021single}, and diffusion of micron-sized tracers in the hydrogels of mucins \cite{cherstvy2019non}. {In this context see also \cite{sabri2020elucidating}, where individual trajectories of quantum dots in the cytoplasm of living cultured cells were described by an intermittent FBM, alternating between two states of different mobility. This research was followed up by \cite{janczura2021identifying}, where two states were modeled by a hidden Markov model and a variant of FBM with $H$ varying between trajectories was studied. We also point the readers' attention to several  experimental works that do not report the consistency with FBM explicitly, however characterize heterogeneity in different cellular environments via random anomalous diffusion exponent \cite{Schwiezer_2015,Pawar_2014,Stiehl_2016,Perkins_2021}.  In particular in Ref. \cite{Stiehl_2016}  the authors 
deduced in this way the heterogeneity of local occupied volume fraction in cellular fluids. Moreover, in Ref. \cite{Perkins_2021} the heterogeneity of anomalous diffusion exponent was analyzed, together with other properties such as skewness of displacement distributions, to discover that the motion of established---i.e. with junction points at each end---endoplasmic reticulum tubules in live Vero cells can be described by sub-populations, and that tubule dynamics are directly related to cellular location.}

{Physically the heterogeneity of the  Hurst exponent and diffusivity discussed above may  correspond to (a) either heterogeneous ensemble of tracers where each tracer has its own Hurst index, or (b) situations where each particle diffuses in its own patch with fixed-within-a-patch but otherwise random diffusivity and random Hurst exponent. Theoretically, such random variations from trajectory to trajectory can be naturally addressed in the framework of superstatistics \cite{s_47,Beck_2005}, which is based on two statistical levels describing, respectively, the fast jiggly dynamics of the diffusing particle and the slow environmental fluctuations with spatially local patches. } We note that this is similar to the idea of the \textit{mixed Poisson process} where the intensity parameter is a random variable \cite{ammeter, book:grandell}.  {In Refs. \cite{beck2021,Itto_2021}} the heterogeneous dynamics of histonelike nucleoid-structuring proteins\cite{wang2018} has been analyzed in the framework of superstatistical approach, where in addition to the fluctuations of the anomalous diffusion exponent due to structural inhomogeneity,  the authors considered the randomness of diffusion coefficient due to temperature fluctuations.

{Even more intricate regimes were discovered by using the advanced deep learning neural network trained on FBM \cite{han2020deciphering}. The authors reported switching in lysosome and endosome movement inside the living eukariotic cells
that can be modeled by FBM with a stochastic Hurst exponent varying in time along a single trajectory. The authors stressed that this was the first application of  }\textit{multifractional process with random Hurst exponent} (MPRE) varying stochastically in time. The MPRE was introduced in Ref. \cite{ayachetaqqu2005} (see also Refs. \cite{bianchi_book,loboda2021}), 
%as a generalization of the \textit{Multifractional fractional Brownian motion} (MFBM),  which stands for FBM with the Hurst eponent being a deterministic function.  \textit{Multifractional Process with Random Exponent} (MPRE)  
where its random wavelet series representation was obtained and process regularity as well as self-similarity were studied. In actuarial mathematics, MPRE corresponds to the idea of \textit{doubly-stochastic Poisson} process (called also the \textit{Cox process}), where the intensity does not only depend on time but is a stochastic process \cite{coxprocess}. {Following Ref. \cite{han2020deciphering}, a comprehensive empirical analysis of the endosomes intracellular motion was provided, which appears to be highly heterogeneous in space and time due to a combination of viscoelasticity, caging, aggregation and active transport \cite{Korabel_arxiv,korabel2021local}. A similar analysis was undertaken in Ref. \cite{Anomalous_Tango} where the authors demonstrated that the motion of Drosophila melanogaster hemocytes in the embrionic phase exhibits a broad spectrum of Hurst exponents and diffusivities, also randomly changing in space and time.} 

{The  experiments and theoretical findings discussed above motivated us to carry out a consistent study of FBM models with a random Hurst exponent. The present paper addresses the FBM model with the Hurst exponent being a random variable (FBMRE). } We provide a general mathematical framework for  the  FBMRE by presenting its probability density function,  moments and the autocovariance function of its increments.  Applying this framework, we present analytical results for three generic distributions of the random Hurst exponent. We discover two remarkable properties of FBMRE, namely accelerating diffusion and persistence transition. We further validate our analytical results with numerical simulations. {In order to highlight the effects emerging due to the randomness of the Hurst exponent we  here omit the case of random diffusivity\cite{Lampo_2017,Witzel_2019} which is supposed to be presented in a forthcoming publication. }Let us also note that several interesting features of FBM with a random diffusion coefficient and constant Hurst exponent have been studied in the framework of diffusing-diffusivity approach in Ref. \cite{wei2020njp}.

The structure of the paper is as follows. In Section \ref{sec:2} we introduce FBMRE and derive general formulas manifesting its basic probabilistic and statistical properties. In Section \ref{sec:3} we consider three special cases of the distributions of the random Hurst exponent, namely two-point, uniform and beta. In Section \ref{sec:4} we present results from numerical simulations which allow to gain insight into accelerating
diffusion and persistence transition phenomena and validate analytical results of Section \ref{sec:3}. Finally, in Section \ref{sec:5} we conclude our analyses by discussing potential relations to experimental data, possible extension of the theory and addressing  the issues which pertain to the representation of spatially heterogeneous processes with spatially independent random parameters.

\section{Fractional Brownian motion with random Hurst exponent}
\label{sec:2}
First, we remind the basic properties of FBM $\{B_{H}(t),~t\geq 0\}$. It is a continuous centered Gaussian process defined through the integral representation \cite{NessMandelbrot} 
%\begin{eqnarray}
%B_{H}(t)=\frac{1}{\Gamma(H+1/2)}\int_{-\infty}^{\infty}\left((t-u)_{+}^{H-1/2}-(-u)_{+}^{H-1/2}\right)dB(u),\nonumber\\
%\end{eqnarray}
\begin{eqnarray}\label{FBM_main}
B_{H}(t)&& =A_{H}\Bigg[\int_{0}^{t}(t-u)^{H-1/2}d\tilde{B}(u)\nonumber\\
&+&\int_{-\infty}^{0} \mspace{-4mu}\left((t-u)^{H-1/2}-(-u)^{H-1/2}\right)d\tilde{B}(u)\Bigg] \mspace{-4mu},
\end{eqnarray}
where $0<H<1$ is the Hurst exponent (called also Hurst index). The process $\{\tilde{B}(t),~t\in \mathbb{R}\}$ is the extension of ordinary Brownian motion  to the negative time axis defined as \begin{equation}
\tilde{B}(t)=    
\begin{cases}
B_1(t) ~ \mbox{for}~t>0,\\
B_2(-t) ~\mbox{for}~ t\leq 0,
\end{cases}
\end{equation}
where $\{B_1(t),~ t \geq 0\}$ and $\{B_2(t),~ t \geq 0\}$ are two independent ordinary Brownian motions. The prefactor $A_{H}=\left[\sqrt{\Gamma\left(2H+1\right)\sin\left(\pi H\right)}/\Gamma\left(H+1/2\right)\right]$ ensures that for given $t\geq 0$, $B_{H}(t)$ is zero-mean Gaussian random variable with variance $\mathbb{E}\left(B_{H}^2(t)\right)=t^{2H}$ (in this paper we use the symbol $\mathbb{E}$ to denote statistical averaging). 
In particular, for $t=1$ the random variable $B_{H}(1)$ has a standard Gaussian distribution, $B_{H}(1)\sim \mathcal{N}(0,1)$. 
Derivation of $A_{H}$ is presented in Appendix \ref{appA}. 

FBM is a self-similar process meaning that   $\{B_H(at)\}$ has the same finite-dimensional distributions as the process $\{a^HB_{H}(t)\}$ for all $a>0$. For $H=1/2$,  FBM becomes an ordinary Brownian motion $\{{B}(t)\}$. 

The increment process of FBM  $\{b_{H}^{\Delta}(t),~t\geq0\}$, which is called fractional Gaussian noise (FGN), is defined as 
\begin{eqnarray}\label{inc}
b_{H}^{\Delta}(t)=B_{H}(t+\Delta)-B_{H}(t),
\end{eqnarray}
where $\Delta$ is a time step.
The FGN is a stationary  process, and its  autocovariance function (ACVF) is given by
\begin{eqnarray}\label{ACVF_FGN}
C_{H}(\tau,\Delta)&=&\mathrm{Cov}(b_{{H}}^{\Delta}(0), b_{{H}}^{\Delta}(\tau))=\mathbb{E}\left(b_{{H}}^{\Delta}(0)b_{{H}}^{\Delta}(\tau)\right)\nonumber\\
&=&\frac{1}{2}\left[(\tau+\Delta)^{2H} + |\tau-\Delta|^{2H} - 2\tau^{2H}\right].
\end{eqnarray}
For small lags $\tau$, such that $\tau/\Delta\ll 1$, one obtains the asymptotic behavior 
\begin{eqnarray}\label{ACVF_FGN_1}
C_{H}(\tau,\Delta)&\sim&\Delta^{2H}\left(1-\left(\frac{\tau}{\Delta}\right)^{2H}\right)
\end{eqnarray}
while for large lags, $\tau/\Delta\gg 1$, one has
\begin{eqnarray}\label{ACVF_FGN_2}
C_{H}(\tau,\Delta)&\sim&\Delta^{2}H(2H-1)\tau^{2H-2}.
\end{eqnarray}
The FGN  has remarkable properties.  For $H>1/2$ it is positively correlated and exhibits the so-called long-range dependence (long-memory or persistence).  In that case the FBM is a superdiffusive process. In contrast, for $H<1/2$ the increment process is negatively correlated and exhibits short-range dependence (anti-persistence), and in that case FBM shows the subdiffusive behaviour.

Now, let us introduce the fractional Brownian motion $\{B_{\mathcal{H}}(t),~t\geq 0\}$ with random Hurst exponent. It is a process defined through the integral representation (\ref{FBM_main}), where the Hurst exponent $H$ is replaced by random variable $\mathcal{H}$ with PDF $f_{\mathcal{H}}(h)$ defined on the interval $(0,1)$ and independent of the process $\{\tilde{B}(t)\}$.

We note that the concept of FBMRE is similar to that of the mixed
Poisson process which plays a prominent role in actuarial
mathematics as a claim counting process. The mixed Poisson process is a generalisation of the classical homogeneous Poisson process with intensity $\lambda>0$, where the intensity parameter is replaced with a positive random variable $\Lambda$ called the structure variable \cite{ammeter,book:grandell}. In actuarial mathematics, the random parameter is attributed to a heterogeneous group of
clients, each one of them generating claims according to a Poisson distribution with the intensity varying from one group to another (for example, in motor insurance to make a
difference between drivers of different age). 
%This idea is not far from the motivation of the fractional Brownian motion with random Hurst exponent.

In this paper we also consider the  increment process $\{b_{\mathcal{H}}^{\Delta}(t),~t\geq 0\}$ of FBMRE, which is defined similar to Eq. (\ref{inc}) with $H$ replaced by $\mathcal{H}$.
%\begin{eqnarray}\label{inc}
%b_{\mathcal{H}}^{\Delta}(t)=B_{\mathcal{H}}(t+\Delta)-B_{\mathcal{H}}(t).
%\end{eqnarray}

In what follows the main attention is paid to the PDF $f_{B_{\mathcal{H}}}(x,t)$ of the process $\{B_{\mathcal{H}}(t)\}$, its MSD $\mathbb{E}\left(B_{\mathcal{H}}^2(t)\right)$, and the ACVF $C_{{\mathcal{H}}}(\tau,\Delta)$ of the process  $\{b_{\mathcal{H}}^{\Delta}(t)\}$. Apparently, as for the  FBM, the increments of $\{B_{\mathcal{H}}(t)\}$ are also stationary.

For given $t>0$ the PDF $f_{B_{\mathcal{H}}}(x,t)$  for $x\in \mathbb{R}$ is given by 
\begin{eqnarray}\label{for1}
    f_{B_{\mathcal{H}}}(x,t) &=& \int_{0}^{1} \frac{1}{\sqrt{2\pi t^{2h}}} \exp\left\{\frac{-x^2}{2t^{2h}} \right\}f_{\mathcal{H}}(h)dh.
    \end{eqnarray}
Eq.~(\ref{for1}) is a direct consequence of the law of total probability \cite{feller} that manifests itself in superstatistical approach to Brownian motion \cite{s_47,Beck_2005}.  

The $q-$th moment  of $B_{\mathcal{H}}(t)$ for any $t>0$ takes the form
\begin{eqnarray}\label{for2}
    \mathbb{E}\left(|B_{\mathcal{{\mathcal{H}}}}(t)|^q\right)&=&c_qM_{\mathcal{H}}(q\log(t)),
\end{eqnarray}
where 
\begin{eqnarray}\label{MGF}
M_{{\mathcal{H}}}(s)=\mathbb{E}(e^{s{\mathcal{H}}})=\int_{0}^{1}e^{sh}f_{\mathcal{H}}(h)dh
\end{eqnarray}
is a moment generating function (MGF) of a random variable ${\mathcal{H}}$ and  
\begin{eqnarray}\label{cq}c_q=\frac{2^{q/2} \Gamma\left(\frac{q+1}{2}\right)}{\sqrt{\pi}}.\end{eqnarray}
In particular, $c_2=1$. 

Let us prove Eq. (\ref{for2}). We first introduce the conditional expectation value $\mathbb{E}(X|Y)$ for any finite-mean random variables $X$ and $Y$ and then remind the law of total expectation \cite{feller}   \begin{eqnarray}
\mathbb{E}\left(X\right)&=&\mathbb{E}\left(\mathbb{E}\left(X| Y\right)\right).
\end{eqnarray}
Then, using the self-similarity property of FBM and the fact that for given fixed ${\mathcal{H}}$,  $\{B_{\mathcal{H}}(t)\}$ is the FBM, we have
\begin{eqnarray*}
  \mathbb{E}\left(|B_{\mathcal{H}}(t)|^q\right)& =& \mathbb{E}\left(\mathbb{E}\left(|B_{{\mathcal{H}}}(t)|^q \Bigg{|} {\mathcal{H}}\right) \right)\\\ &=& \mathbb{E}\left( \mathbb{E}\left(t^{q{\mathcal{H}}} |B_{{\mathcal{H}}}(1)|^q\Bigg|{\mathcal{H}}\right) \right).
  \end{eqnarray*}
Let us recall that for given fixed $\mathcal{H}$ the random variable  $B_{{\mathcal{H}}}(1)$ has a standard Gaussian distribution. Taking the notation $Z=|B_{{\mathcal{H}}}(1)|$ we obtain 
\begin{eqnarray*}
\mathbb{E}\left( \mathbb{E}\left(t^{q{\mathcal{H}}} Z^q|{\mathcal{H}}\right) \right)
        &=&
        \mathbb{E}\left(Z^q \mathbb{E}\left(t^{q{\mathcal{H}}}|{\mathcal{H}}\right) \right) \\&=& 
        \mathbb{E}\left(Z^q t^{q{\mathcal{H}}}\right) = \mathbb{E}\left(Z^q \right)  \mathbb{E}(t^{q{\mathcal{H}}})\\
        &=&\mathbb{E}\left(Z^q \right)  \mathbb{E}(e^{{q{\mathcal{H}}}\log(t)})\\&=&  \mathbb{E}\left(Z^q \right)M_{\mathcal{H}}(q \log(t)).
    \end{eqnarray*}
It is easy to see that $\mathbb{E}\left(Z^q \right)$ is given by Eq.~(\ref{cq}) and thus, we arrive at Eq. (\ref{for2}).

Taking $q=2$, one can calculate the  MSD of $\{B_{\mathcal{H}}(t)\}$,
\begin{eqnarray}\label{for3}
    \mathbb{E}\left(B_{\mathcal{H}}^2(t)\right)=M_{\mathcal{H}}(2\log(t)).
\end{eqnarray}
Using the form of the ACVF of the increment process of FBM (see Eq.~(\ref{ACVF_FGN})), we obtian the ACVF of  $\{b_{\mathcal{H}}^{\Delta}(t)\}$,\begin{eqnarray}\label{for4}
    C_{{\mathcal{H}}}(\tau,\Delta)&=& \mathrm{Cov}(b_{\mathcal{H}}^{\Delta}(0), b_{\mathcal{H}}^{\Delta}(\tau))\nonumber\\
   &=& \mathbb{E}\left(b_{\mathcal{H}}^{\Delta}(0)  b_{\mathcal{H}}^{\Delta}(\tau)\right) = \mathbb{E}\left(\mathbb{E}(b_{{H}}^{\Delta}(0) b_{{H}}^{\Delta}(\tau)|{\mathcal{H}}=H ) \right)\nonumber  \\
     &=&  \frac{1}{2}\mathbb{E}\left((\tau+\Delta)^{2{\mathcal{H}}} + |\tau-\Delta|^{2{\mathcal{H}}} - 2\tau^{2{\mathcal{H}}}\right)\nonumber\\&=&
     \frac{1}{2} \mathbb{E}\left(e^{2{\mathcal{H}}\log(\tau+\Delta)} + e^{2{\mathcal{H}}\log|\tau-\Delta|} - 2e^{2{\mathcal{H}}\log(\tau)} \right)\nonumber\\&=&
     \frac{1}{2}\left[ M_{\mathcal{H}}\left(2\log(\tau+\Delta)\right) +M_{\mathcal{H}}\left(2\log|\tau-\Delta|\right)\right]\nonumber \\&-& M_{\mathcal{H}}\left(2\log{ \tau}\right).
\end{eqnarray}
In the next section, we consider  special cases of the distribution for the random variable ${\mathcal{H}}$. 

\section{Generic distributions of random Hurst exponent}
\label{sec:3}
\subsection{Two-point distribution of the random Hurst exponent}\label{sectionA}
Let us first consider the simplest example of the {random} Hurst exponent, namely the variable ${\mathcal{H}}$ that has a {(univariate) probability distribution concentrated in two points}   $H_1,H_2\in (0,1)$, $H_1<H_2$ ,
    \begin{eqnarray}\label{two_pdf}f_{\mathcal{H}}(h)=
    p\delta(h-H_1)+(1-p)\delta(h-H_2),
    \end{eqnarray}
where  $\delta(\cdot)$ is a Dirac delta function while $p$ and $1-p$ are the corresponding probability masses, { $p\in (0,1)$}.    {For clarity, we call the PDF (\ref{two_pdf}) as two-point distribution which obviously should not be mixed with bivariate probability density.}
    
The MGF of ${\mathcal{H}}$, Eq. (\ref{MGF}), is given by 
\begin{eqnarray}\label{MGF_two}
M_{\mathcal{H}}(s)=pe^{H_1s}+(1-p)e^{H_2s}.\end{eqnarray}
Using Eqs. (\ref{for1}) and (\ref{two_pdf}) we obtain the PDF of $B_{\mathcal{H}}(t)$,
\begin{eqnarray}\label{main5}
    f_{B_{\mathcal{H}}}(x,t) &=& \frac{p}{\sqrt{2\pi t^{2H_1}}} \exp\left\{\frac{-x^2}{2t^{2H_1}} \right\}\nonumber\\&+&\frac{1-p}{\sqrt{2\pi t^{2H_2}}} \exp\left\{\frac{-x^2}{2t^{2H_2}} \right\}.
    \end{eqnarray}
From here we conclude that   $B_{\mathcal{H}}(t)$ has a distribution that is a mixture of Gaussian distributions with zero-means and variances equal to $t^{2H_1}$ and $t^{2H_2}$. The probability  to stay at the origin reads
\begin{eqnarray}\label{origin}
    f_{B_{\mathcal{H}}}(0,t) &=& \frac{p}{\sqrt{2\pi t^{2H_1}}} + \frac{1-p}{\sqrt{2\pi t^{2H_2}}} ,
    \end{eqnarray}
and thus it demonstrates the decay $\sim t^{-H}$ determined by the larger exponent $H=H_2$ at short times, $t\ll1$, and smaller exponent $H=H_1$ at long times, $t\gg1$, respectively. The MSD takes the form
  \begin{eqnarray}\label{for17}
  \mathbb{E}\left(B_{\mathcal{H}}^2(t)\right)&=&\left[pt^{2H_1}+(1-p)t^{2H_2}\right],
\end{eqnarray}
and shows the growth $\sim t^{2H}$ determined by the smaller exponent $H = H_1$ at short times ($t\ll1$) and  by larger exponent $H = H_2$ at long times ($t\gg1$), respectively. Following the terminology used in the theory of distributed order  fractional diffusion equations, we may call this effect as \textit{accelerating diffusion} \cite{PhysRevE.66.046129,PhysRevE.78.021111,doi:10.1142/9789814340595_0005}.  In what follows we will demonstrate the universality of this effect for other distributions of the random Hurst exponent. This phenomenon is the first main result of the paper.

%A. V. Chechkin, R. Gorenflo, I. M. Sokolov, Retarding subdiffusion and accelerating superdiffusion governed by distributed order fractional diffusion equations. Physical Review E 66, No.046129, 1-7 (2002).\\
%A.V. Chechkin, V.Yu. Gonchar, R. Gorenflo, N. Korabel, and I.M. Sokolov,
%Generalized fractional diffusion equations for accelerating subdiffusion and truncated Lévy flights, Physical Review E 78, Article 021111(1-13) (2008).\\
%A. Chechkin, I.M. Sokolov, J. Klafter, Natural and Modified Forms of Distributed-Order Fractional Diffusion Equations. In: J. Klafter, S.C. Lim, R. Metzler (Eds), Fractional Dynamics: Recent Advances. Singapore: World Scientific, 2011. Chapter 5, pp. 107-127.}.

Finally, using Eqs.~(\ref{for4}) and (\ref{MGF_two}) we obtain the ACVF for the increment process 
\begin{eqnarray}\label{twopointsacvf}
    C_{{\mathcal{H}}}(\tau,\Delta)&=&pC_{H_1}(\tau,\Delta)+(1-p)C_{H_2}(\tau,\Delta),
    %\frac{p}{2}(\tau+\Delta)^{2H_1}+\frac{1-p}{2}(\tau+\Delta)^{2H_2}\nonumber\\&+&\frac{p}{2}|\tau-\Delta|^{2H_1}+\frac{1-p}{2}|\tau-\Delta|^{2H_2}\nonumber\\&-&p\tau^{2H_1}-(1-p)\tau^{2H_2}.
      \end{eqnarray}
      where $C_H(\tau,\Delta)$ is given by Eq. (\ref{ACVF_FGN}).
      Making an expansion of Eq. (\ref{twopointsacvf}) for two limit cases of small and large $\tau/\Delta$  we get  the  asymptotic formulas
      \begin{eqnarray}\label{twopointsacvf1}
      C_{{\mathcal{H}}}(\tau,\Delta)&\sim& p\Delta^{2H_1}\left(1-\left(\frac{\tau}{\Delta}\right)^{2H_1}\right)\nonumber\\
      &+&(1-p)\Delta^{2H_2}\left(1-\left(\frac{\tau}{\Delta}\right)^{2H_2}\right),
           \end{eqnarray}
          for $\tau/\Delta\ll 1$,  and  
      \begin{eqnarray}\label{eq20}C_{{\mathcal{H}}}(\tau,\Delta)&\sim& p\Delta^2H_1(2H_1-1)\tau^{2(H_1-1)}\nonumber\\&+&(1-p)\Delta^{2}H_2(2H_2-1)\tau^{2(H_2-1)},\end{eqnarray}
     for $\tau/\Delta \gg 1$, respectively.
One can see that at very long times the decay of the ACVF is determined by the second term in Eq. (\ref{eq20}). However in intermediate time scale the first term can be dominant. If $H_1<1/2<H_2$, then an interesting effect can be observed, namely, the transition from the antipersistent to persistent behavior.   We will see in what follows that such transition also occurs for other generic examples considered in this paper. Such  transition from antipersistent and persistent regimes, which we call \textit{persistence transition}, is the second main result of our paper.
%\textcolor{red}{one can conclude that $C_{{\mathcal{H}}}(\tau,\Delta)$ tends to zero asymptotically as $|\tau|^{2(H_2-1)}$. Add the discussion}

\subsection{Uniform distribution of the random Hurst exponent}\label{section_unif}
As the second example we consider the case when  the Hurst exponent has a uniform distribution on the interval $[H_1,H_2]$, $0< H_1<H_2<1$. In this case the PDF and MGF are given  by
    \begin{eqnarray}\label{unif_pdf}f_{\mathcal{H}}(h)=\frac{{I}_{h\in [H_1,H_2]}}{H_2-H_1},~~
   M_{\mathcal{H}}(s)=\frac{e^{H_2s}-e^{H_1s}}{s(H_2-H_1)},\end{eqnarray}
respectively, where ${I}_{h\in [H_1,H_2]}$ is  equal to $1$ when \\$H_1\leq h\leq H_2$ and $0$ otherwise.  Using Eq.~(\ref{for1})  we obtain 
\begin{eqnarray}\label{pdf_unif}
        f_{B_{\mathcal{H}}}(x,t) &= &\frac{1}{H_2-H_1} \int_{H_1}^{H_2} \frac{\exp\left\{-x^2/(2 t^{2h}) \right\}}{\sqrt{2\pi t^{2h}}}   dh.
        \end{eqnarray}
Making substitution $y=x/t^{h}$ we obtain 
\begin{eqnarray}\label{pdf_unifo}
        f_{B_{\mathcal{H}}}(x,t) &=&\frac{\Phi\left({x}/{t^{H_1}}\right)-\Phi\left({x}/{t^{H_2}}\right)}{x(H_2-H_1)\log (t)},
    \end{eqnarray}
 where $\Phi(\cdot)$ is the cumulative distribution function of the standard Gaussian distribution,
 \begin{eqnarray}
 \Phi(x)=\int_{-\infty}^{x}\frac{\exp\left\{-u^2/2 \right\}}{\sqrt{2\pi }}du.
 \end{eqnarray}
The probability to stay at the origin is obtained from Eq. (\ref{pdf_unif}),
\begin{eqnarray}
       f_{B_{\mathcal{H}}}(0,t)=\frac{1}{\sqrt{2\pi}(H_2-H_1)\log(t)}\left(\frac{1}{t^{H_1}}-\frac{1}{t^{H_2}}\right)\mspace{-4mu},
\end{eqnarray}
and thus, similar to the case A of the two-point distribution, it is determined by the largest Hurst index $H_2$ for short times and smallest Hurst index $H_1$ for long times. 
The large $x$ asymptotic behavior is obtained from Eq. (\ref{pdf_unifo}) using the asymptotics  
\begin{eqnarray}\label{asym2}
 \Phi(x)\sim 1-\frac{e^{-x^2/2}}{\sqrt{2\pi}x},
\end{eqnarray}
and thus the asymptotic shape of PDF is determined by smallest Hurst index $H_1$ for short times, 
\begin{eqnarray}
f_{B_{\mathcal{H}}}(x,t)\sim\frac{t^{H_1}e^{-x^2/2t^{2H_1}}}{\sqrt{2\pi}(H_2-H_1)x^2\log (1/t)},~t\ll 1,
\end{eqnarray}
while the largest Hurst index $H_2$ determines the asymptotic shape for long times,
 \begin{eqnarray}
        f_{B_{\mathcal{H}}}(x,t) &\sim&\frac{t^{H_2}e^{-x^2/2t^{2H_2}}}{\sqrt{2\pi}(H_2-H_1)x^2\log (t)},~t\gg 1.
    \end{eqnarray}
Using Eq. (\ref{for3}) we obtain the formula for MSD 
\begin{eqnarray}\label{for29}
%\mathbb{E}(B_{\mathcal{H}}^q(t))&=&c_q\left[\frac{t^{bq}-t^{aq}}{q(b-a)\log (t)}\right],\\
             \mathbb{E}(B_{\mathcal{H}}^2(t)) &=& \frac{t^{2H_2} - t^{2H_1}}{2(H_2-H_1)\log(t)}.
\end{eqnarray} 
This formula gives us the simple asymptotics for short times 
\begin{eqnarray}\label{for30}
\mathbb{E}(B_{\mathcal{H}}^2(t)) &\sim& \frac{t^{2H_1}}{2(H_2-H_1)\log(1/t)}, ~~t\ll 1
\end{eqnarray}
and long times
\begin{eqnarray}\label{for31}
\mathbb{E}(B_{\mathcal{H}}^2(t)) &\sim& \frac{t^{2H_2}}{2(H_2-H_1)\log(t)}, ~~t\gg 1,
\end{eqnarray}
and thus we again observe the phenomenon of accelerating diffusion. 
%similar to the case of two-point distribution $f_{\mathcal{H}}(h)$ in Case A. 

Finally, using (\ref{for4}) and the MGF given in (\ref{unif_pdf}) we obtain  the ACVF for $\{b_{\mathcal{H}}^{\Delta}(t)\}$
 \begin{eqnarray}\label{acvf_unif}C_{{\mathcal{H}}}(\tau,\Delta)     &=&\frac{1}{2(H_2-H_1)}\left(\frac{(\tau+\Delta)^{2H_2}}{\log(\tau+\Delta)}+\frac{|\tau-\Delta|^{2H_2}}{\log(|\tau-\Delta|)}\right)\nonumber\\
 &-&\frac{1}{2(H_2-H_1)}\left(\frac{(\tau+\Delta)^{2H_1}}{\log(\tau+\Delta)}+\frac{|\tau-\Delta|^{2H_1}}{\log(|\tau-\Delta|)}\right)\nonumber\\
     &-&\frac{1}{(H_2-H_1)}\left(\frac{\tau^{2H_2}}{\log(\tau)}-\frac{\tau^{2H_1}}{\log(\tau)}\right).\end{eqnarray}
The asymptotics of ACVF for $\tau/\Delta \ll 1$ and $\tau/\Delta \gg 1$ are demonstrated in Appendix \ref{app_unif}. However, the persistence transition can not be easily seen from the analytical formulas there. This phenomenon is further demonstrated for a more general case of beta distribution in numerical simulations discussed in Section  \ref{sec:4}. 
%Thus, one can conclude that for large $\tau$ $\gamma_{b_{\mathcal{H}}^{\Delta}}(\tau)$ behaves asymptotically as $|\tau|^{2(b-1)}/\log(\tau)$.
\subsection{Beta distribution on the interval $[H_1,H_2]$ of the random Hurst exponent}
As the last and the most important example, we consider the case when the Hurst exponent has a beta distribution on the interval $[H_1,H_2]$, $0<H_1<H_2<1 $ with parameters $\alpha>0$ and $\beta>0$ (see Eq. (\ref{beta_pdf}) for the definition). The choice of beta distribution is motivated {at least by its three} advantages: first, it is defined on the finite support; second,  the parameters $\alpha$ and $\beta$ provide an essential flexibility in shape; {and third, the beta distribution is amenable to analytical studies}.  Let us note that the uniform distribution on the interval $[H_1,H_2]$ is a special case of the beta distribution on the same interval with the parameters $\alpha=\beta=1$ and thus all the results obtained in this section can be reduced to those obtained in Section \ref{section_unif}. 
   
For the readers' convenience in Appendix \ref{app1} we present some results for the particular case of  the beta distributed Hurst exponent on the interval $(0,1)$.  Now at first we note that the beta distributed random variable $\mathcal{H}$ defined on the interval $[H_1,H_2]$ is expressed via beta distributed random variable $\mathcal{H}_1$ on the interval $(0,1)$ as
%To this end we employ the PDF and MGF given  in Appendix \ref{app1}, Eqs. (\ref{beta_pdf}) and (\ref{beta_mgf}), for the Beta distribution on the interval $(0,1)$. At first note that a 
   \begin{eqnarray}
   \mathcal{H}=(H_2-H_1)\mathcal{H}_1+H_1.
   \end{eqnarray}
Thus, the PDF of $\mathcal{H}$ is given by
\begin{eqnarray}f_{\mathcal{H}}(h)&=&\frac{1}{H_2-H_1}f_{\mathcal{H}_1}\left(\frac{h-H_1}{H_2-H_1}\right),
\end{eqnarray}
where $f_{\mathcal{H}_1}(h)$ is the PDF given in Eq. (\ref{beta_pdf}).
Therefore,
\begin{eqnarray}
f_{\mathcal{H}}(h)&=&\frac{(h-H_1)^{\alpha-1}(H_2-h)^{\beta-1}}{\mathbb{B}(\alpha,\beta)(H_2-H_1)^{\alpha+\beta-1}}{I}_{h\in [H_1,H_2]},
\end{eqnarray}
where 
\begin{eqnarray}\label{beta_fun}\mathbb{B}(\alpha,\beta)=\frac{\Gamma(\alpha)\Gamma(\beta)}{\Gamma(\alpha+\beta)}\end{eqnarray}
is a beta function.

The MFG of ${\mathcal{H}}$ is given by
   \begin{eqnarray}\label{beta_mgf_new}
      M_{\mathcal{H}}(s) = e^{sH_1}M_{\mathcal{H}_1}(s(H_2-H_1)),
         \end{eqnarray}
where the $M_{\mathcal{H}_1}(s)$ is a MGF given Eq. (\ref{beta_mgf}).
Therefore,
\begin{eqnarray}\label{mgf_beta_h1h2}
M_{\mathcal{H}}(s)=e^{sH_1}{}_1 F_1(\alpha, \alpha+\beta, s(H_2-H_1)),
\end{eqnarray}
where ${}_1 F_1(\cdot, \cdot, \cdot)$ is a confluent hypergeometric function defined in Eq. (\ref{F1function}) \cite{abramowitz}.

Using Eq.~(\ref{for1}) we obtain the PDF, 
% \begin{eqnarray}
%  f_{B_{\mathcal{H}}}(x,t) &=& \frac{1}{(H_2-H_1)^{\alpha+\beta-1}\mathbb{B}(\alpha,\beta)}\times\nonumber\\
%  &&\int_{H_1}^{H_2} \frac{(h-H_1)^{\alpha-1}(H_2-h)^{\beta-1}}{\sqrt{2\pi t^{2h}}} \exp\left\{\frac{-x^2}{2t^{2h}} \right\}dh.
% \end{eqnarray}
\begin{eqnarray}
 f_{B_{\mathcal{H}}}(x,t) &=& \int_{H_1}^{H_2} \frac{(h-H_1)^{\alpha-1}(H_2-h)^{\beta-1}\exp\left\{\frac{-x^2}{2t^{2h}} \right\}}{\sqrt{2\pi t^{2h}}} dh\nonumber\\
 &\times& \frac{1}{(H_2-H_1)^{\alpha+\beta-1}\mathbb{B}(\alpha,\beta)}.
\end{eqnarray}
By the use of Eq. (\ref{for2}) the MSD takes the form
    \begin{eqnarray}    \label{for40} \mathbb{E}\left(B_{\mathcal{H}}^2(t)\right)&=&t^{2H_1}\times\nonumber\\&&{}_1 F_1(\alpha, \alpha+\beta, 2(H_2-H_1)\log(t)),
\end{eqnarray}
which apparently reduces to Eq.~(\ref{app1_msd}) for $H_1=0$ and $H_2=1$.

Using Eq.~(\ref{special_function_small}) and the asymptotics Eq.~(\ref{sim1}) we obtain the asymptotics for MSD at short times
\begin{eqnarray}\label{beta_small}
\mathbb{E}\left(B_{\mathcal{H}}^2(t)\right)&\sim&
\frac{\Gamma(\alpha+\beta)t^{2H_1}}{\Gamma(\alpha)(2(H_2-H_1)\log(1/t))^{\alpha}}.
\end{eqnarray}

The asymptotics for confluent hypergeometric function, Eq.~(\ref{sim1}), gives the asymptotics for MSD at long times,
\begin{eqnarray}\label{beta_large}
\mathbb{E}\left(B_{\mathcal{H}}^2(t)\right)&\sim&\frac{\Gamma(\alpha+\beta)t^{2H_2}}{\Gamma(\alpha)(2(H_2-H_1)\log(t))^{\beta}}.
\end{eqnarray}
Note that Eqs.~(\ref{beta_small})  and (\ref{beta_large}) reduce to Eqs.~(\ref{beta_large_1}) and (\ref{beta_small_1}) for $H_2=1$ and $H_1=0$.
Finally,  Eqs. (\ref{for4}) and (\ref{mgf_beta_h1h2})  give the ACVF for $\{b_{\mathcal{H}}^{\Delta}(t)\}$,
%following the lines of Appendix \ref{app1}
% \begin{eqnarray}
% C_{{\mathcal{H}}}(\tau,\Delta) &=&\frac{1}{2} (\tau+\Delta)^{2H_1}{}_1 F_1(\alpha, \alpha+\beta,2(H_2-H_1)\log(\tau+\Delta))\nonumber\\&+&\frac{1}{2}|\tau-\Delta|^{2H_1}{}_1 F_1(\alpha, \alpha+\beta,2(H_2-H_1)\log(| \tau-\Delta|))\nonumber\\
% &-& \tau^{2H_1}{}_1 F_1(\alpha, \alpha+\beta,2(H_2-H_1)\log(\tau)).\end{eqnarray}
\begin{widetext}
\begin{eqnarray}\label{acvf_beta_h1h2}
C_{{\mathcal{H}}}(\tau,\Delta) &=& \frac{1}{2} \Big\{(\tau+\Delta)^{2H_1} \Big.
{}_1 F_1\left(\alpha, \alpha+\beta,2(H_2-H_1)\log(\tau+\Delta)\right) +  |\tau-\Delta|^{2H_1} {}_1 F_1\left(\alpha, \alpha+\beta,2(H_2-H_1)\log| \tau-\Delta|\right)  \nonumber\\
&-& \Big. 2\tau^{2H_1}{}_1 F_1\left(\alpha, \alpha+\beta,2(H_2-H_1)\log\tau\right) \Big\},\end{eqnarray}
\end{widetext}
which reduces to the particular a case for $H_1=0$, $H_2=1$,  as it should be (see Eq.~(\ref{acvf_beta_zero_one})).
The above expression for the ACVF can be simplified in the limits of short and long time lags $\tau$ (see Appendix \ref{app22}), however the existence of the persistence transition is not so obvious as it is seen  from  simple Eqs.~(\ref{twopointsacvf}), (\ref{twopointsacvf1}) and (\ref{eq20}) for two-point distribution of the Hurst exponent.
To study the  persistence transition in the next section numerically we employ the exact formula given by Eq.~(\ref{acvf_beta_h1h2}) for particular cases of the parameters.
%The asymptotics for ACVF are demonstrated in Appendix 
 %\ref{app2}. The persistence transition phenomenon for the considered case is discussed in the simulation section.

{To conclude this section we would like to stress that using model distributions like two-point and uniform ones is convenient if one strives to demonstrate new effects in a most “pure” form.  Apparently, the general methodology provided in this paper can be applied to any distribution on the interval $[0,1]$. For example, we also performed analytical and numerical calculations for asymmetric triangular distribution, and the results are qualitatively similar to those presented in the paper. } 
\section{Numerical analysis}

\label{sec:4}
In this section we perform numerical analyses of different characteristics of the FBMRE that will help us to discover important and distinct features of the process.
We analyse the behaviour of MSD and PDF for FBMRE, and ACVF of its increments by using three different choices of the Hurst exponent distribution, namely two-point, uniform and beta distributions which were studied analytically in detail in Section \ref{sec:3}. 
To this end, we simulate trajectories of the process and employ the Monte Carlo method. Since the FBMRE can be viewed as a result of a two-step randomization procedure, first, the Hurst exponent is generated and second, the FBM with that generated value is simulated

\begin{figure*}[ht] 
  \centering
  \includegraphics[width=.7\linewidth]{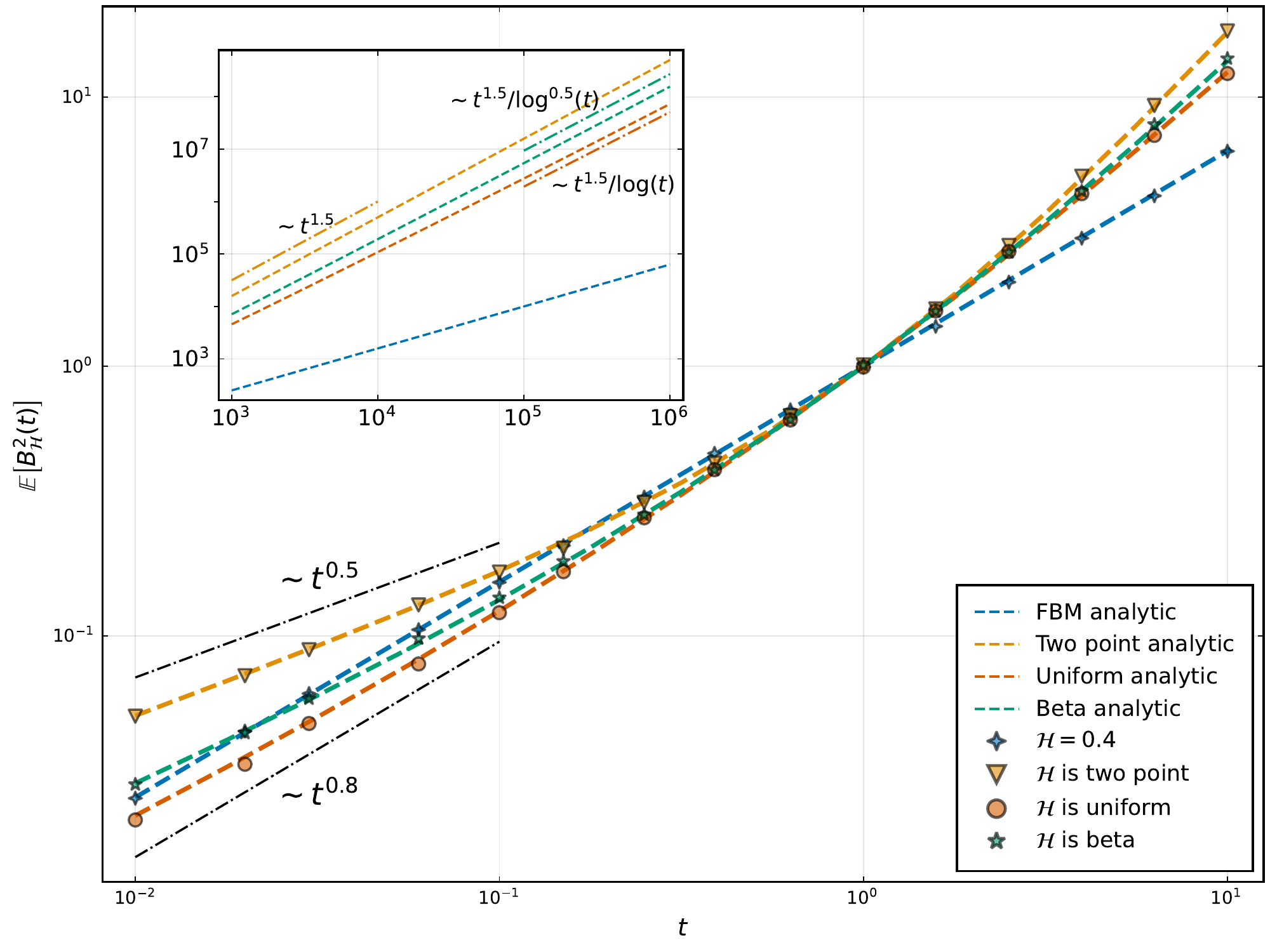}  
  \caption{Mean squared displacement of FBMREs with two-point, uniform and beta distributions of the Hurst exponent calculated analytically by using the results from Section \ref{sec:3} (different lines), and by means of Monte Carlo simulations (different markers). For the two-point distribution $H_1=0.25$, $H=0.75$ with $p=0.5$,  uniform is distributed on $(0.25,0.75)$, and beta is distributed on (0.25,0.75) with $\alpha=\beta = 0.5$. The reference blue line and blue diamond markers correspond to FBM with $H=0.4$. The inset depicts the analytical behaviour for longer times and the asymptotics are plotted next to the relevant cases. }
  \label{fig:msd}
\end{figure*}

In Figure  \ref{fig:msd} we present the MSD  calculated both analytically (see Eqs. (\ref{for17}), (\ref{for29}) and (\ref{for40})) and by means of Monte Carlo simulations. For the latter case, we generated $10,000$ trajectories of FBMRE for the time interval $[0,10]$ with $\Delta=0.01$.
The parameters of the two-point distribution are $H_1=0.25$, $H_2=0.75$ and $p=0.5$, uniform distribution is concentrated on the interval $(0.25, 0.75)$, and beta distribution is defined on the interval $(0.25,0.75)$ with the parameters $\alpha=\beta = 0.5$.
The parameters of the distributions are chosen in such a way to make them close to one another, namely they have the same range and mean.

The most important observation is that there is a transition from subdiffusion at shorter times to superdiffusion at longer times  observed for all considered FBMREs: the diffusion is clearly accelerating.
We also observe that the analytical results agree well with simulations. Moreover, the results for short and long times match the asymptotic formulas presented for all three types of distributions considered, see Eqs. (\ref{for17}) (\ref{for30}), (\ref{for31}), (\ref{beta_small})  and (\ref{beta_large}).  

\begin{figure}
    \centering
    \includegraphics[width=.9\linewidth]{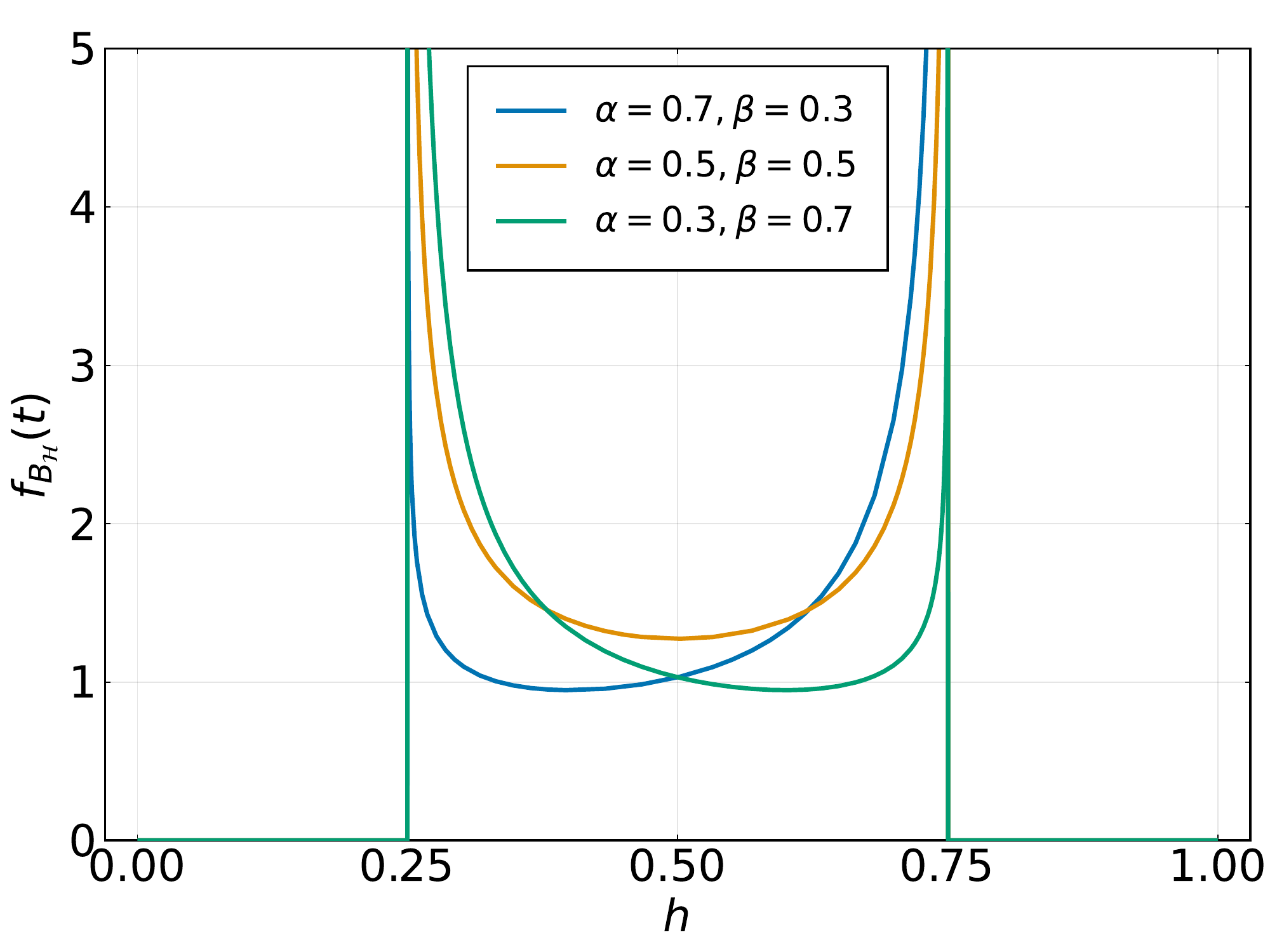}
    \caption{Probability density function for the beta distribution defined on the interval $(0.25,0.75)$ for three different pairs of the parameters: $\alpha=0.7$ and $\beta=0.3$ (blue line); $\alpha=0.5$ and $\beta=0.5$ (orange line); $\alpha=0.3$ and $\beta=0.7$ (green line).   }
    \label{fig:beta_density}
\end{figure}

We now concentrate on two distributions, which we believe, can serve as the most important generic cases, namely the two-point and beta. The two-point distribution has the probability masses $p$ and $1-p$ at $H_1=0.25$ and $H_2=0.75$, respectively.
The beta distribution is defined on the same range $(0.25,0.75)$. Further, for the two-point distribution we analyse three cases, namely $p=0.1$ ('almost' superdiffusion), $p=0.5$ (intermediate case), and $p=0.9$ ('almost' subdiffusion). 
Similarly, for the beta distribution we select three pairs of parameters $\alpha=0.7$, $\beta=0.3$; $\alpha=0.5$. $\beta=0.5$, and $\alpha=0.3$, $\beta=0.7$. The PDFs for the three selected cases are presented in Figure \ref{fig:beta_density}. From here one can see that for $\alpha=0.7$ and $\beta=0.3$ the distribution is skewed to the right (the same type of skewness as for the two-point distribution with $p=0.1$); for $\alpha=0.5$ and $\beta=0.5$ the distribution is symmetric around its mean  $H=0.5$ (similar to two-point distribution  with $p=0.5$), and  
%the largest part of the probability is concentrated to the right of the central point
for $\alpha=0.3$ and $\beta=0.7$ the distribution is skewed to the left (as for the two-point distribution with $p=0.9$).
%the largest part of the probability is concentrated to the left side of the central point 
%while in case . We thus may expect that these three cases exhibit features that are similar to the three types of two-point distributions with $p=0.5$, $p=0.9$ and $p=0.1$, respectively.

\begin{figure*} % pdf
\begin{subfigure}{.45\textwidth}
  \centering
  % include first image
  \includegraphics[width=.8\linewidth]{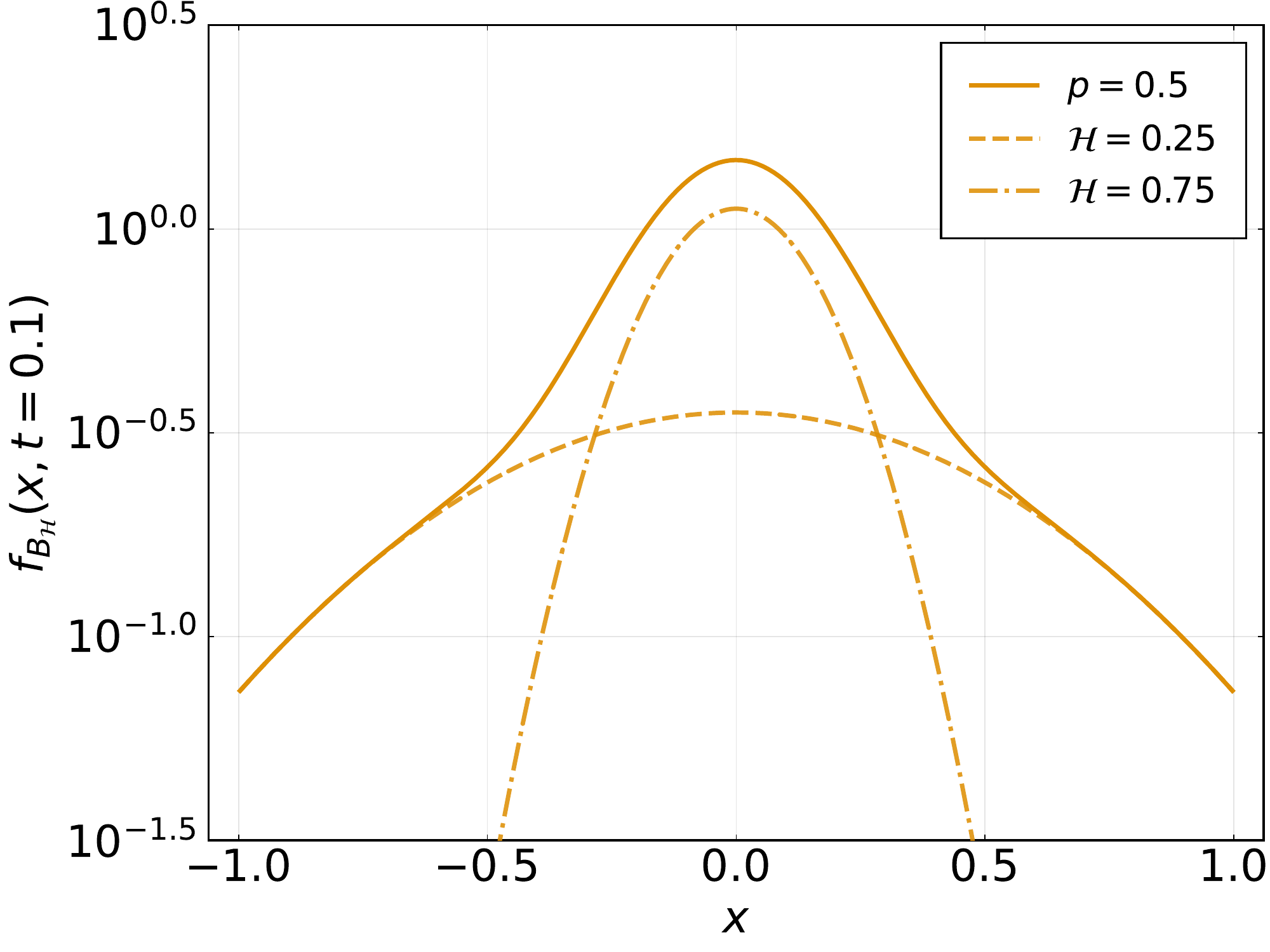}  
  \vskip -.2cm
  \caption{}
  \label{fig:pdf-2point-short}
\end{subfigure}
\begin{subfigure}{.45\textwidth}
  \centering
  % include second image
  \includegraphics[width=.8\linewidth]{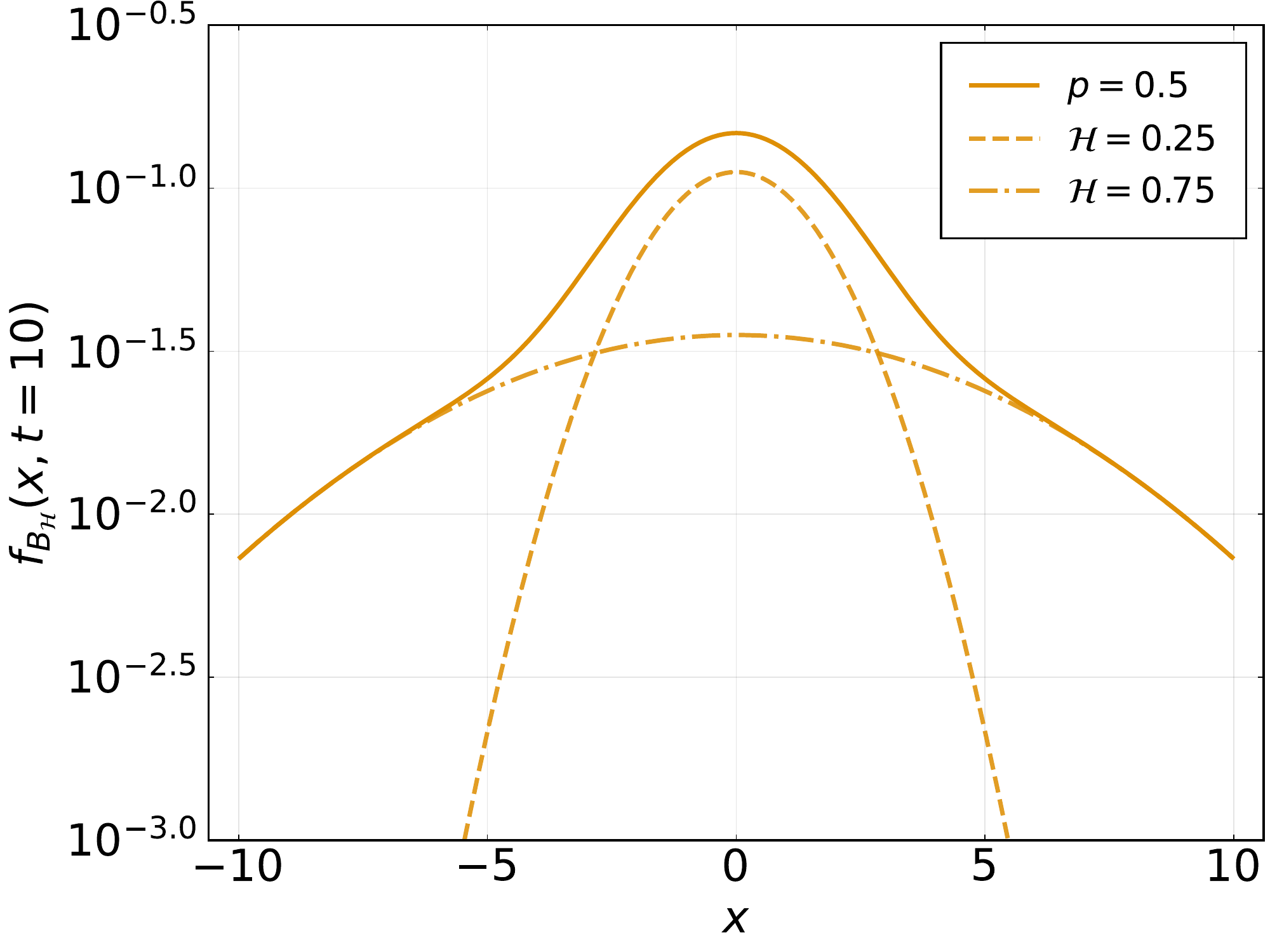}
  \vskip -.2cm
  \caption{}
  \label{fig:pdf-2point-long}
\end{subfigure}
\\
\begin{subfigure}{.45\textwidth}
  \centering
  % include second image
  \includegraphics[width=.8\linewidth]{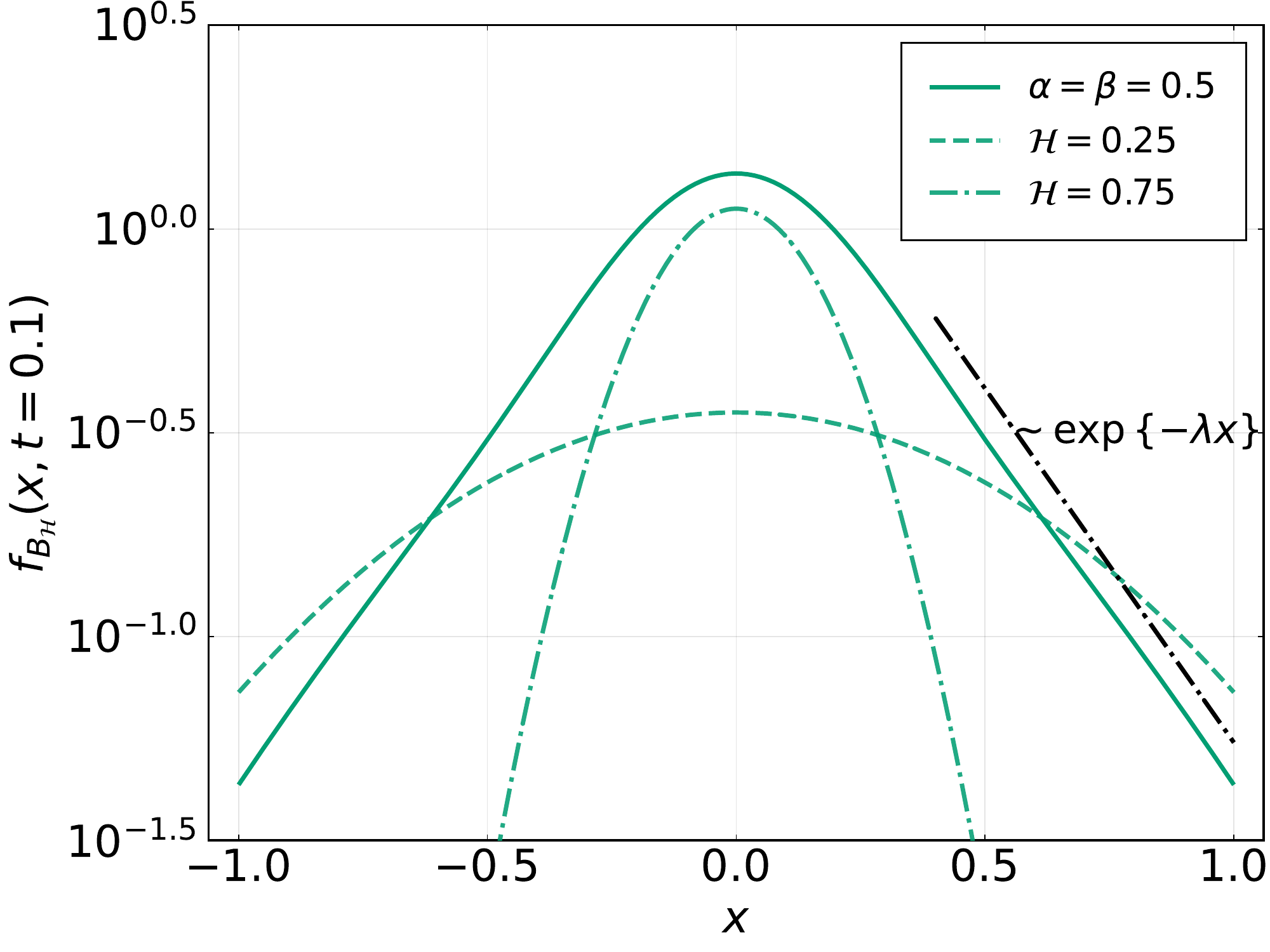}
  \vskip -.2cm
  \caption{}
  \label{fig:pdf-beta-short}
\end{subfigure}
\begin{subfigure}{.45\textwidth}
  \centering
  % include second image
  \includegraphics[width=.8\linewidth]{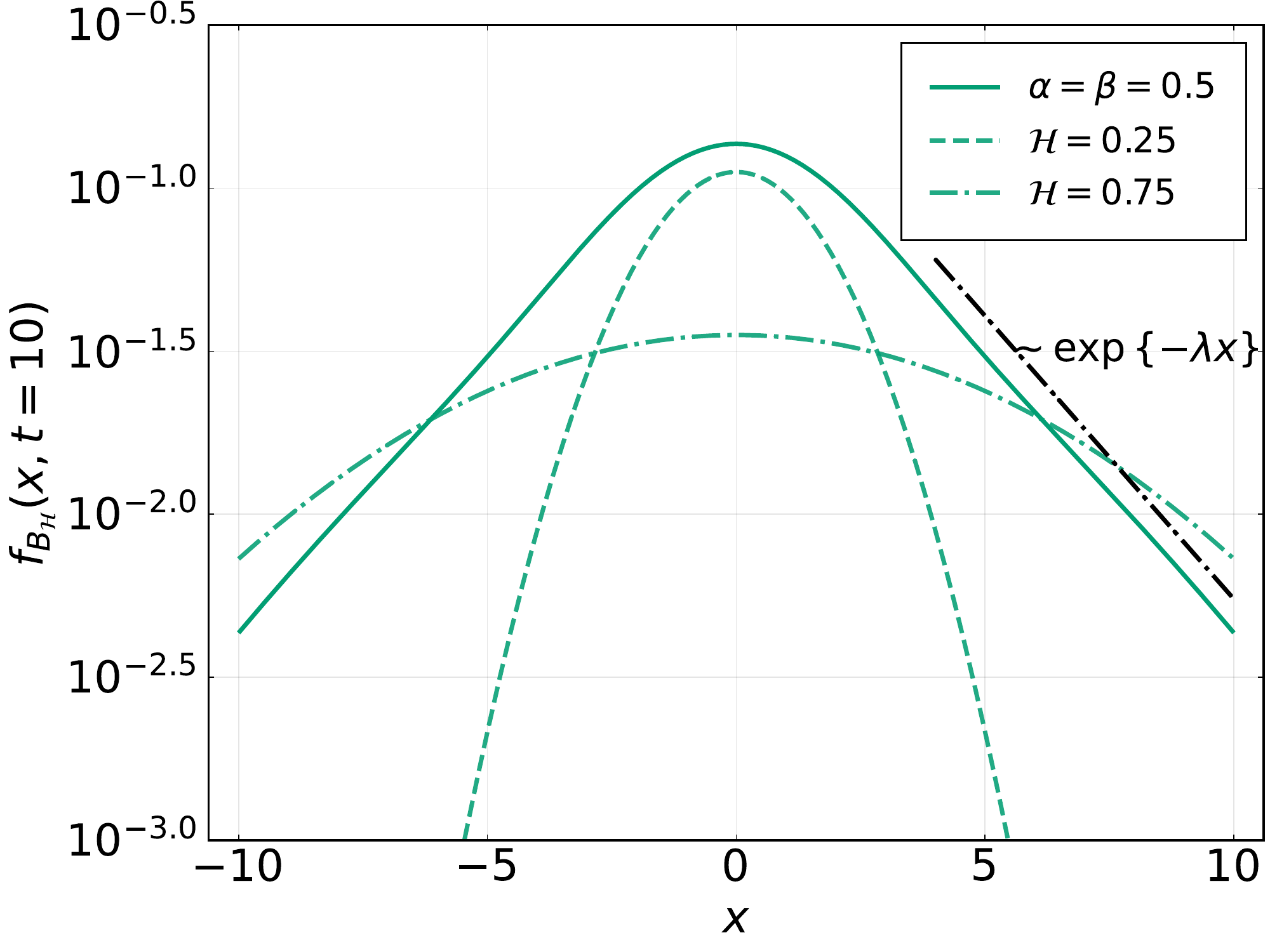}
  \vskip -.2cm
  \caption{}
  \label{fig:pdf-beta-long}
\end{subfigure}

\caption{
Probability density function of FBMREs for two-point (top panels) and beta (bottom panels) distributions of the Hurst parameter for two time points: $t=1$ (left panels) and $t=10$ (right panels) (solid lines). The parameters of the distributions are the same as used in Figure \ref{fig:msd}. Dashed green lines correspond to FBM with $H=0.25$ and dot-dashed green lines to FBM with $H=0.75$. Dot-dashed black lines represent exponential decay of the tails of the distribution, with $\lambda=4$ (bottom left panel) and $\lambda = 0.4$ (bottom right panel).}
\label{fig:pdf}
\end{figure*}

In Figure \ref{fig:pdf} we depict PDFs of the FBMREs for the two distributions of the Hurst parameter, and PDFs of the FBM for $H=0.25$ and $H=0.75$, for two time points: $t=1$ and $t=10$. The parameters of the distributions are the same as used in Figure \ref{fig:msd} for the intermediate case $p=0.5$.
In Figures \ref{fig:pdf}(a)-(b) the behaviour of the two-point distribution is shown. We observe that for short time $t=1$ the center of the distribution resembles the PDF of FBM with $H=0.75$, while the tails coincide with those of FBM with $H=0.25$. For long time $t=10$ situation is quite opposite: the tails correspond to FBM with $H=0.75$ and the distribution center to FBM with $H=0.25$. In Figures \ref{fig:pdf}(c)-(d) the beta distribution case is presented. The most striking difference from the two-point distribution case is the fact the tails are no longer close to Gaussian and they rather resemble an exponential decay. 
%\textcolor{red}{AW and ACH: add the sentence when Michal will add the lines on the figure}. 
This is due to the fact that for the two-point distribution, the FBMRE follows a mixture of two normal laws, whereas for the beta distribution  the resulting distribution structure is more complex and leads to exponential-like tails.

\begin{figure*} % acvf
\begin{subfigure}{.45\textwidth}
  \centering
  % include first image
  \includegraphics[width=.8\linewidth]{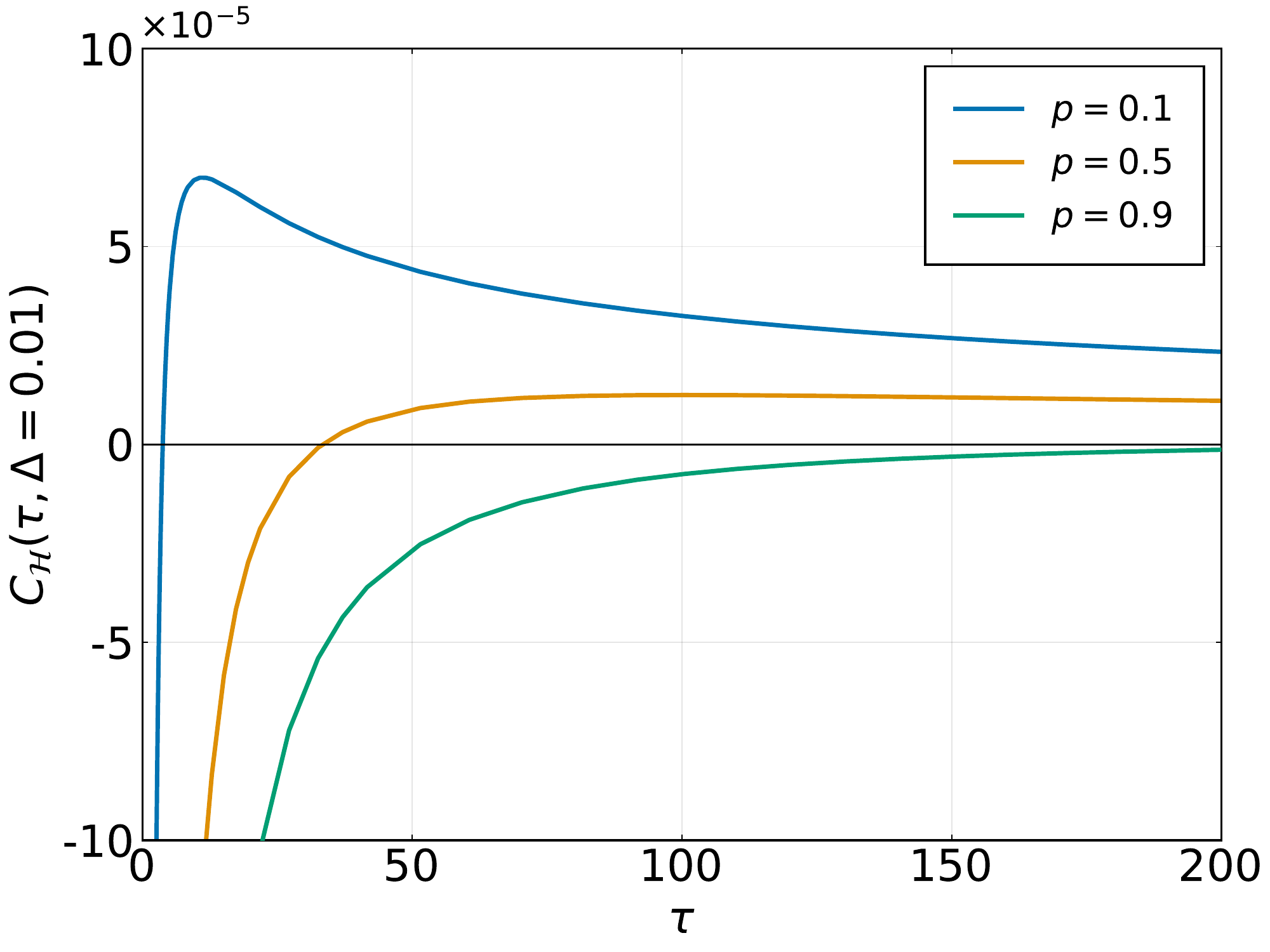}  
  \vskip -.2cm
  \caption{}
  \label{fig:acvf-2point}
\end{subfigure}
\begin{subfigure}{.45\textwidth}
  \centering
  % include second image
  \includegraphics[width=.8\linewidth]{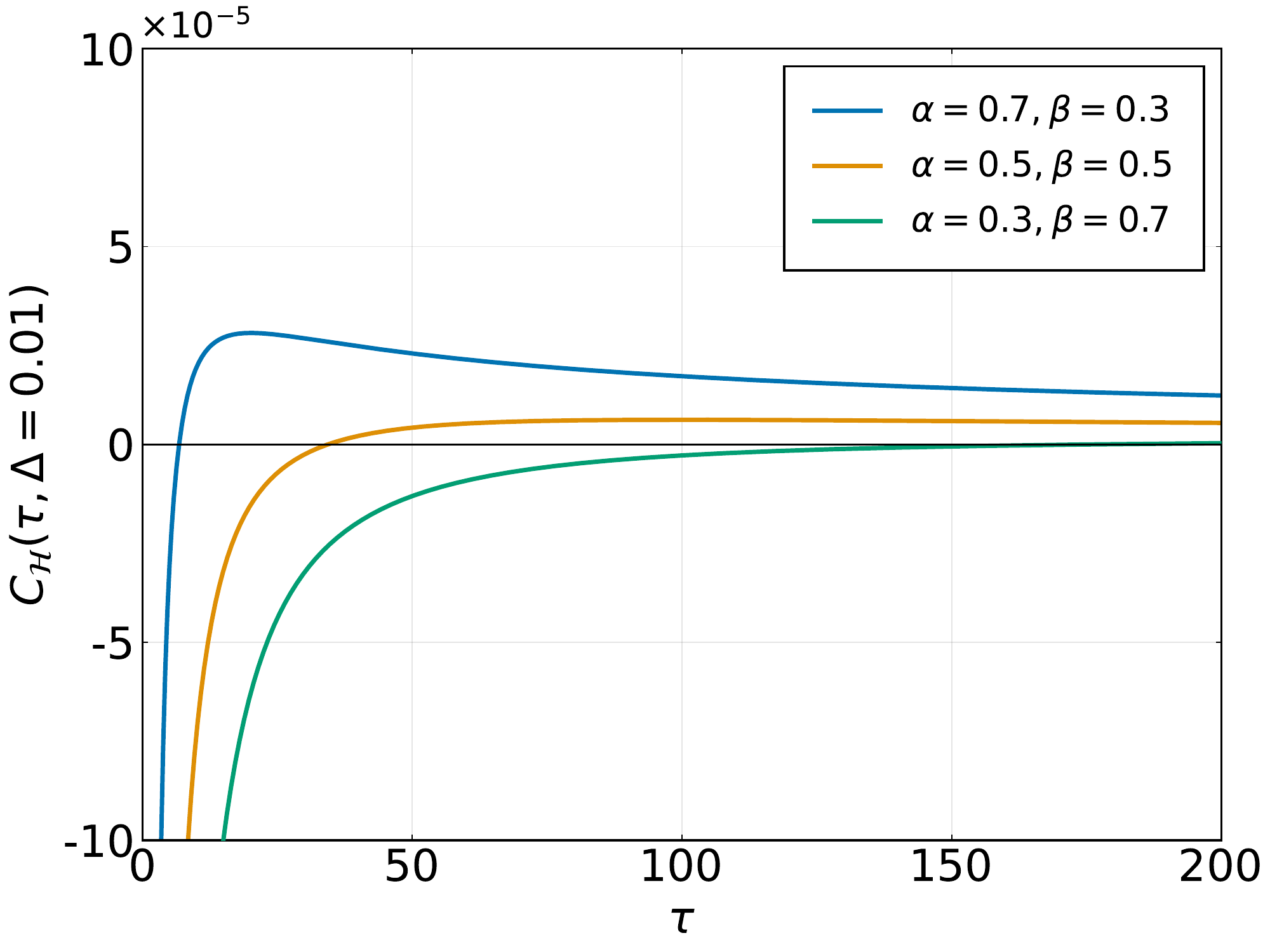}
  \vskip -.2cm
  \caption{}
  \label{fig:acvf-beta}
\end{subfigure}

\caption{Long-time ACVF $C_{\mathcal{H}}(\tau,\Delta)$  of the FBMRE increments for $\Delta=0.01$ with two-point (panel (a)) and beta (panel (b)) distributions of the Hurst exponent. For the two-point distribution the parameters are  $H_1=0.25$, $H_2=0.75$, and $p=0.1$ (blue line), $0.5$ (orange line), $0.9$ (green line).  The beta distribution is defined on $(0.25,0.75)$ with parameters $\alpha=0.7$, $\beta=0.3$ (blue line), $\alpha=0.5$, $\beta=0.5$ (orange line), and $\alpha=0.3$, $\beta=0.7$ (green line). For clarity of the presentation we do not plot the ACVF for lag equal to zero which is the variance of the increment process. }
\label{fig:acvf}
\end{figure*}

In Figure \ref{fig:acvf}, the long-time behaviour of the  ACVFs for the two distributions is depicted as the functions of the lag $\tau$. For the two-point distribution, see Figure \ref{fig:acvf}(a), we can see that the probability masses assigned to the points $H_1=0.25$, $H_2=0.75$ have a significant impact on the behaviour of ACVF, namely if the mass assigned to $H_1=0.25$ is small ($p=0.1$), then the ACVF is positive (apart from the very small lags), which corresponds to the superdiffusive (persistent) case. In contrast, for $p=0.9$ the ACVF is negative and converges to zero which matches the subdiffusive (antipersistent) case. For the probability mass $p=0.5$, the situation is intermediate, namely the ACVF starts with negative values, at lag around $30$ ACVF crosses the zero level and then it is positive, hence the process clearly performs the persistence transition. A similar behaviour can be also observed in  Figure \ref{fig:acvf}(b) for the beta distribution defined on the interval 
$(0.25,0.75)$ for three corresponding choices of the distribution parameters. When the parameters $\alpha$ and $\beta$ are equal then the persistence transition is again clearly visible. 

\begin{figure*} % carpet cov
\begin{subfigure}{.3\textwidth}
  \centering
  % include first image
  \includegraphics[width=.8\linewidth]{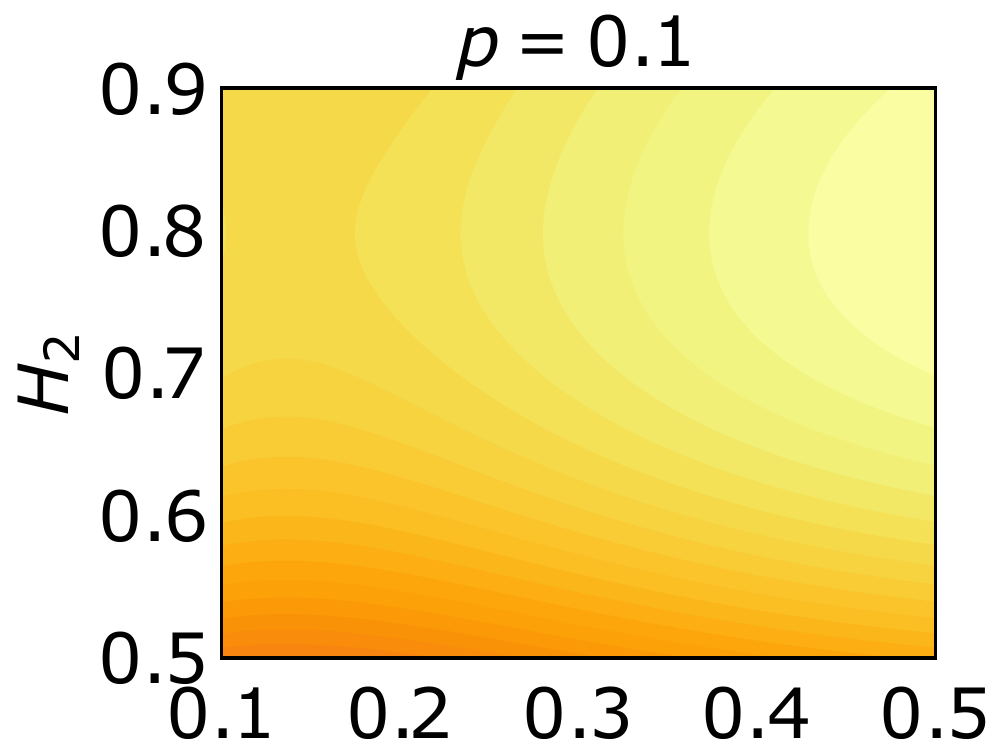}  
  \vskip -.2cm
  \caption{}
  \label{fig:a1_low}
\end{subfigure}
\begin{subfigure}{.3\textwidth}
  \centering
  % include second image
  \includegraphics[width=.8\linewidth]{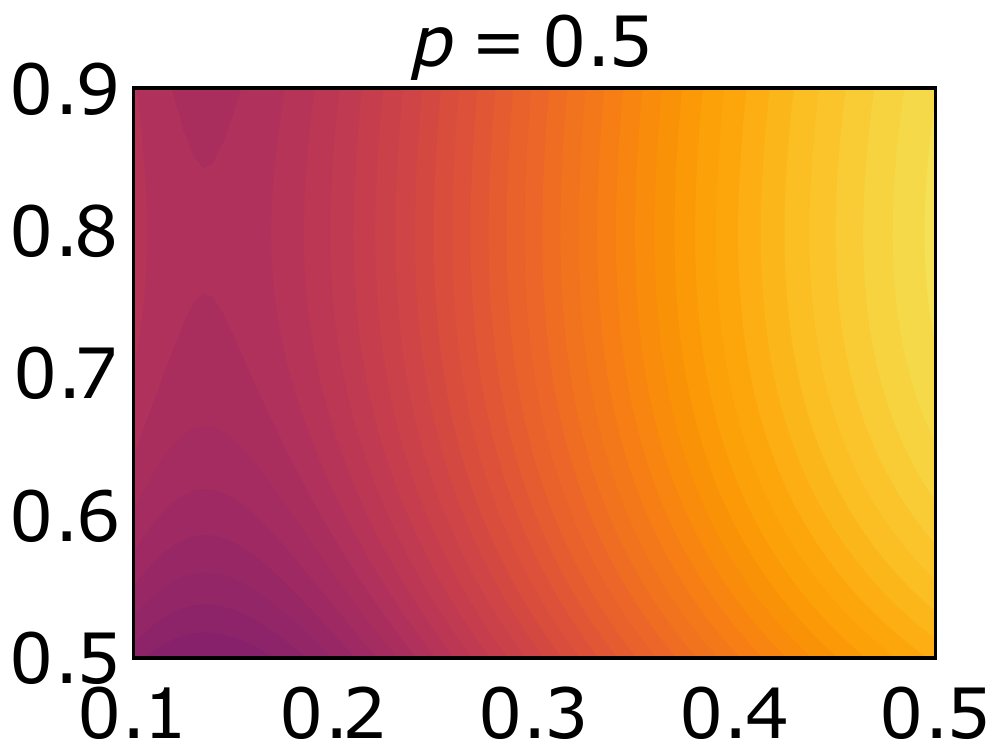}
  \vskip -.2cm
  \caption{}
  \label{fig:a2_low}
\end{subfigure}
\begin{subfigure}{.3\textwidth}
  \centering
  % include second image
  \includegraphics[width=.9\linewidth]{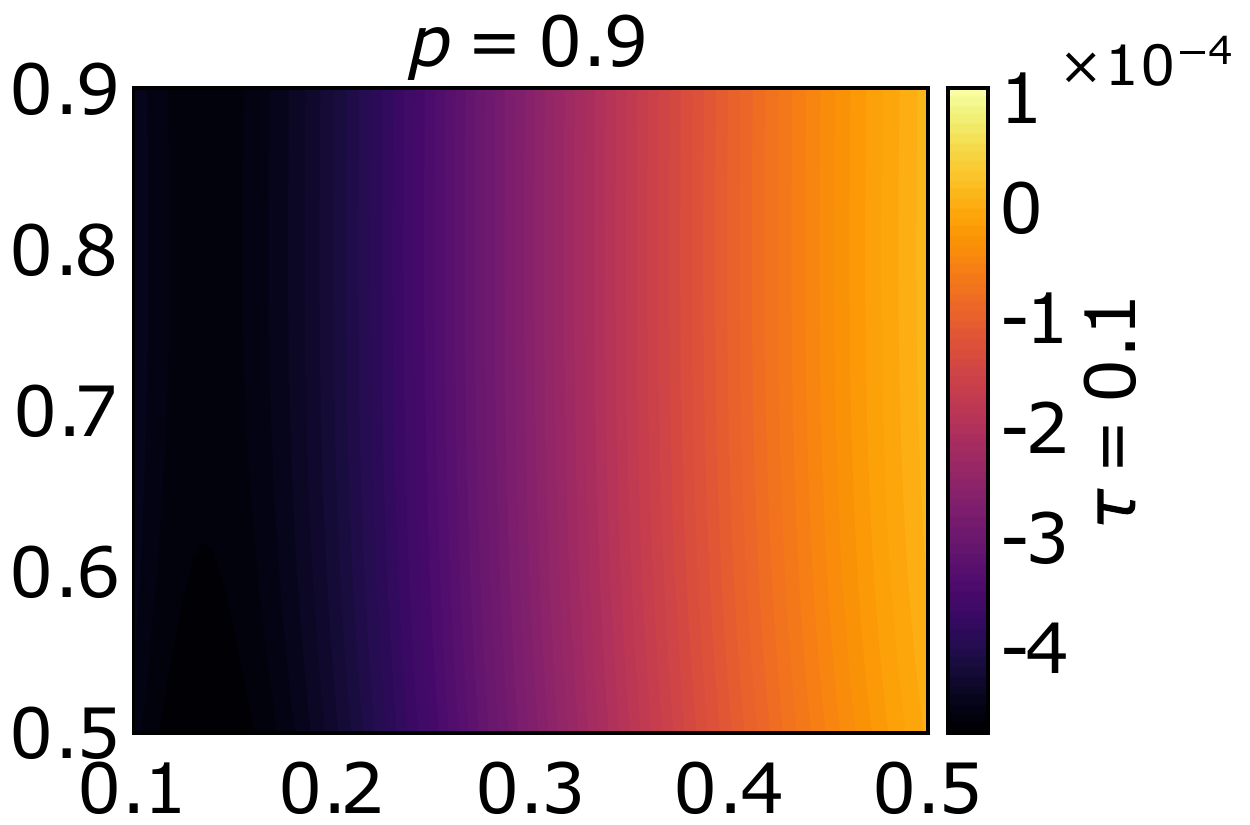}
  \vskip -.2cm
  \caption{}
  \label{fig:a3_low}
\end{subfigure}
\\

\begin{subfigure}{.3\textwidth}
  \centering
  % include first image
  \includegraphics[width=.8\linewidth]{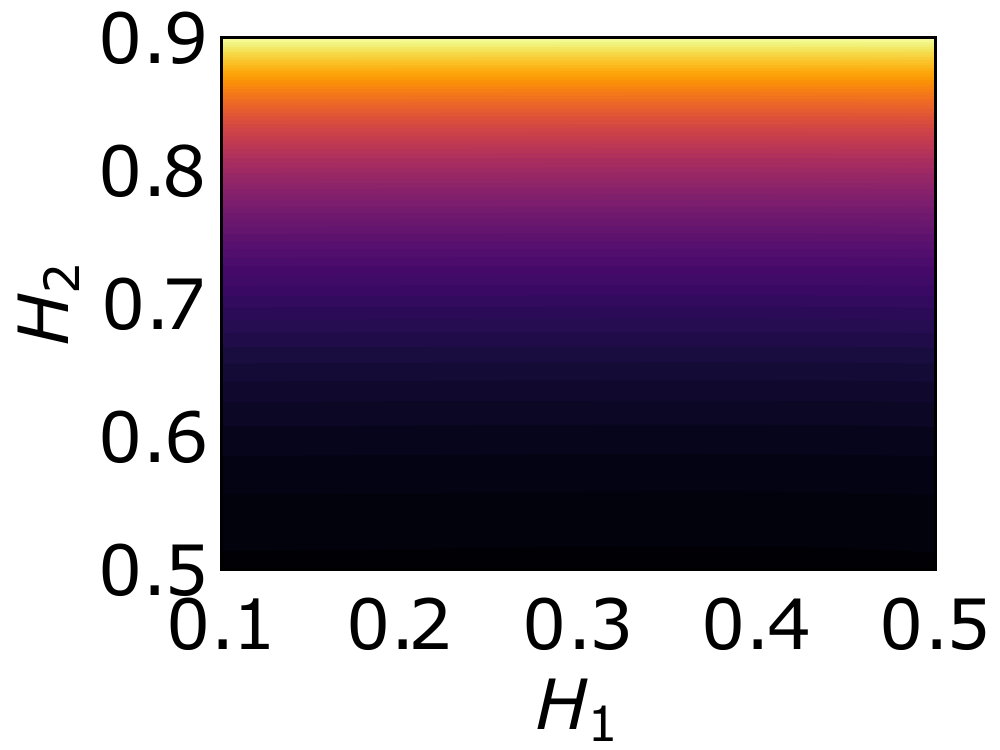}  
  \vskip -.2cm
  \caption{}
  \label{fig:a1_high}
\end{subfigure}
\begin{subfigure}{.3\textwidth}
  \centering
  % include second image
  \includegraphics[width=.8\linewidth]{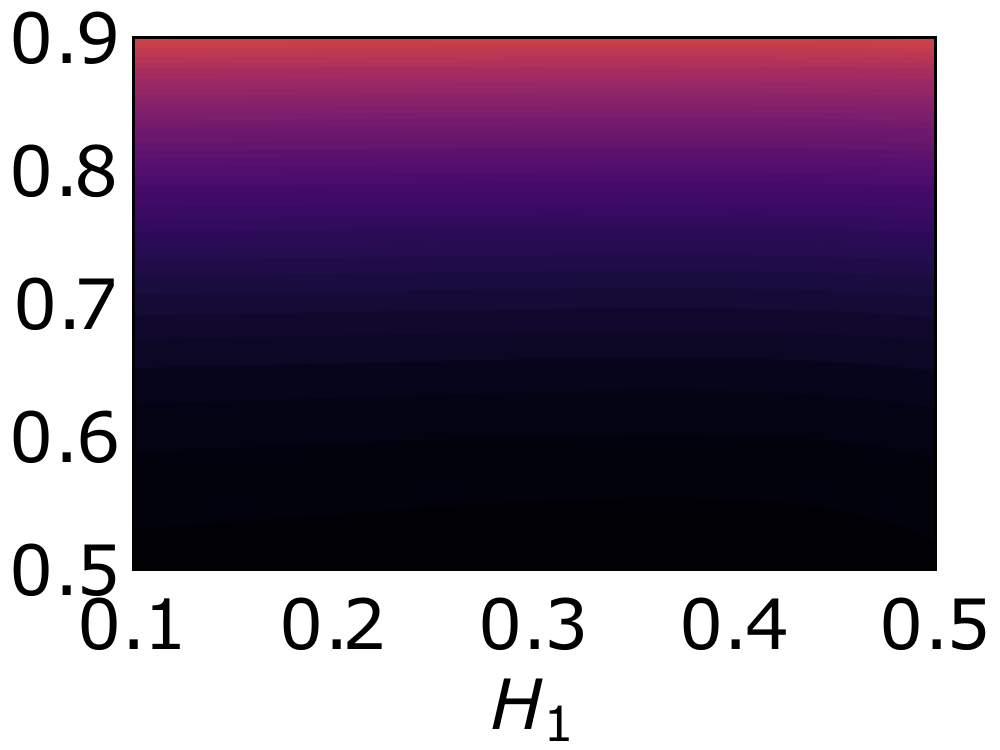}
  \vskip -.2cm
  \caption{}
  \label{fig:a2_high}
\end{subfigure}
\begin{subfigure}{.3\textwidth}
  \centering
  % include second image
  \includegraphics[width=.9\linewidth]{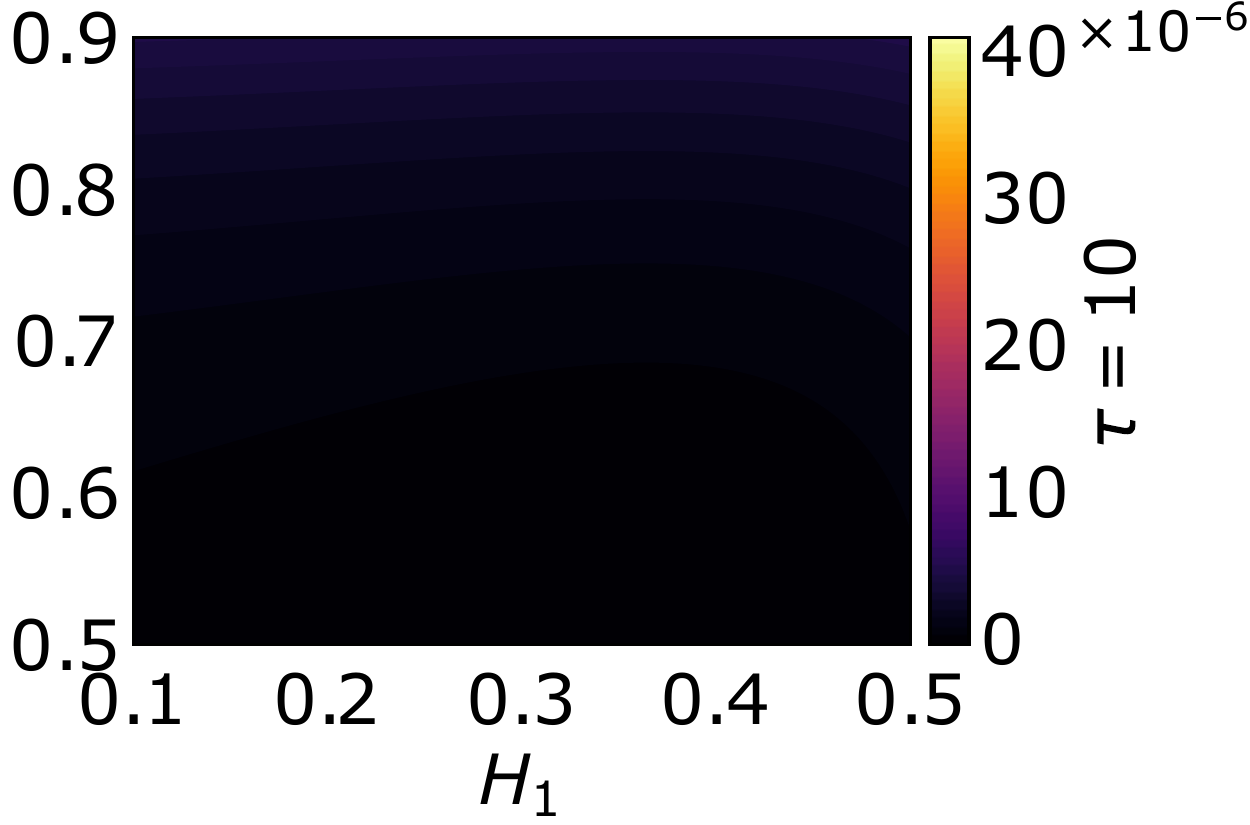}
  \vskip -.2cm
  \caption{}
  \label{fig:a3_high}
\end{subfigure}
\caption{Carpet plots of ACVFs for the FBMRE increments for $\Delta=0.01$ with the two-point distribution of the Hurst exponent defined on the interval $(H_1,H_2)$ for the three cases of parameter $p=0.1$ (left panels), $0.5$ (middle panels), and $0.9$ (right panels), and for two different lags $\tau=0.1$ (top panels) and $\tau=10$ (bottom panels).}
\label{fig:acvf_carpet}
\end{figure*}

\begin{figure*} % carpet cov beta
\begin{subfigure}{.3\textwidth}
  \centering
  % include first image
  \includegraphics[width=.8\linewidth]{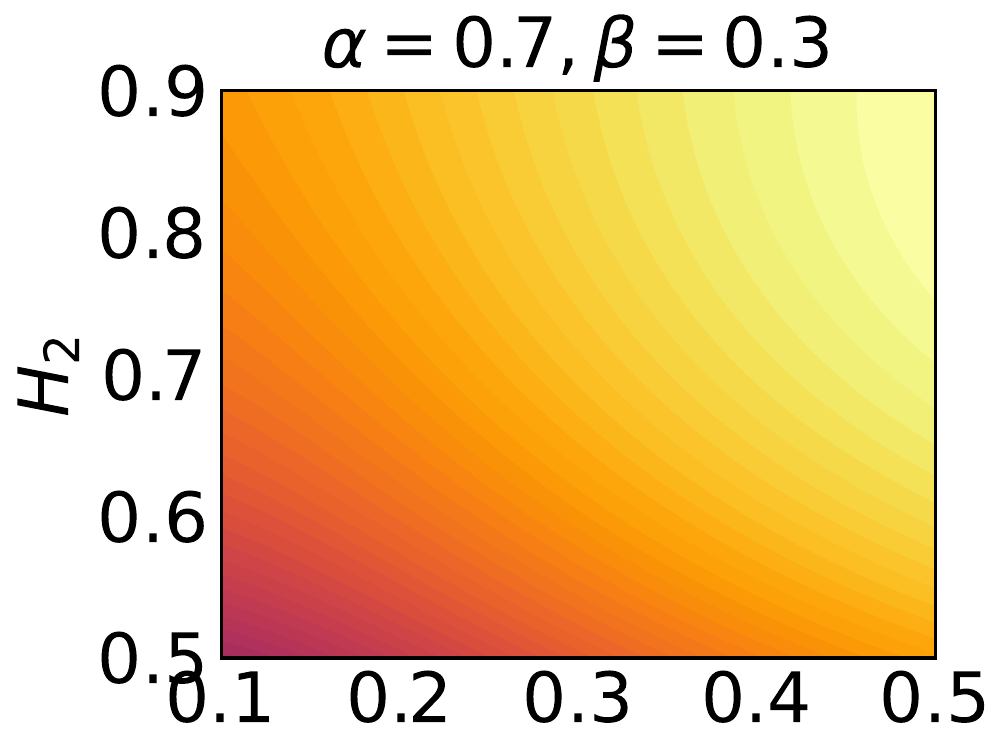}  
  \vskip -.2cm
  \caption{}
  \label{fig:c1_low}
\end{subfigure}
\begin{subfigure}{.3\textwidth}
  \centering
  % include second image
  \includegraphics[width=.8\linewidth]{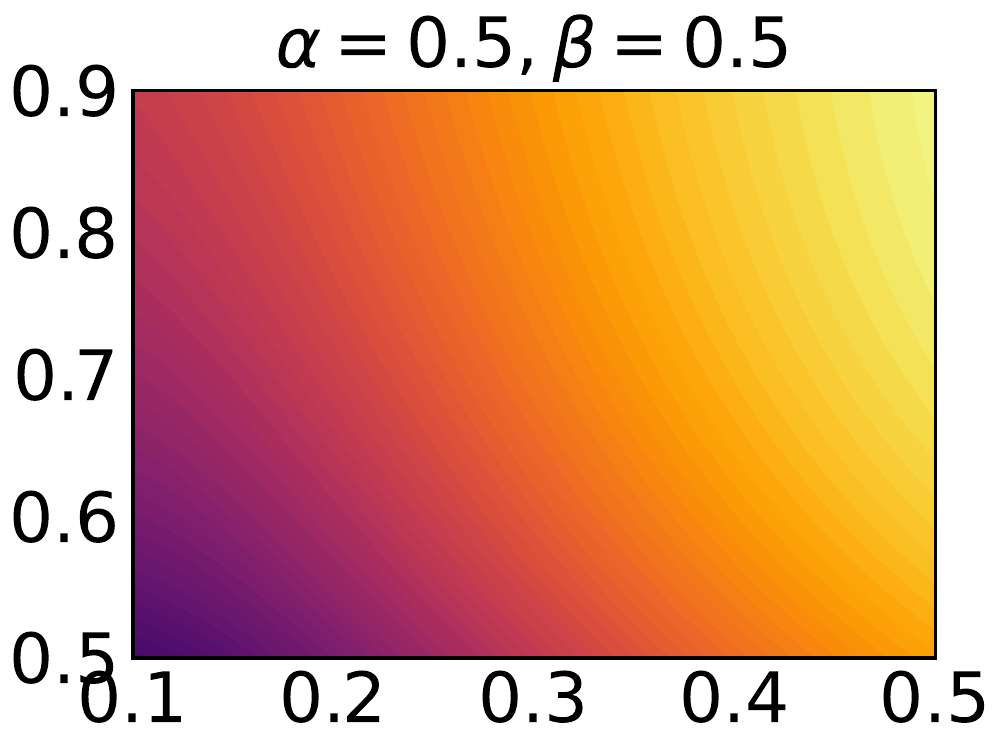}
  \vskip -.2cm
  \caption{}
  \label{fig:c2_low}
\end{subfigure}
\begin{subfigure}{.3\textwidth}
  \centering
  % include second image
  \includegraphics[width=.9\linewidth]{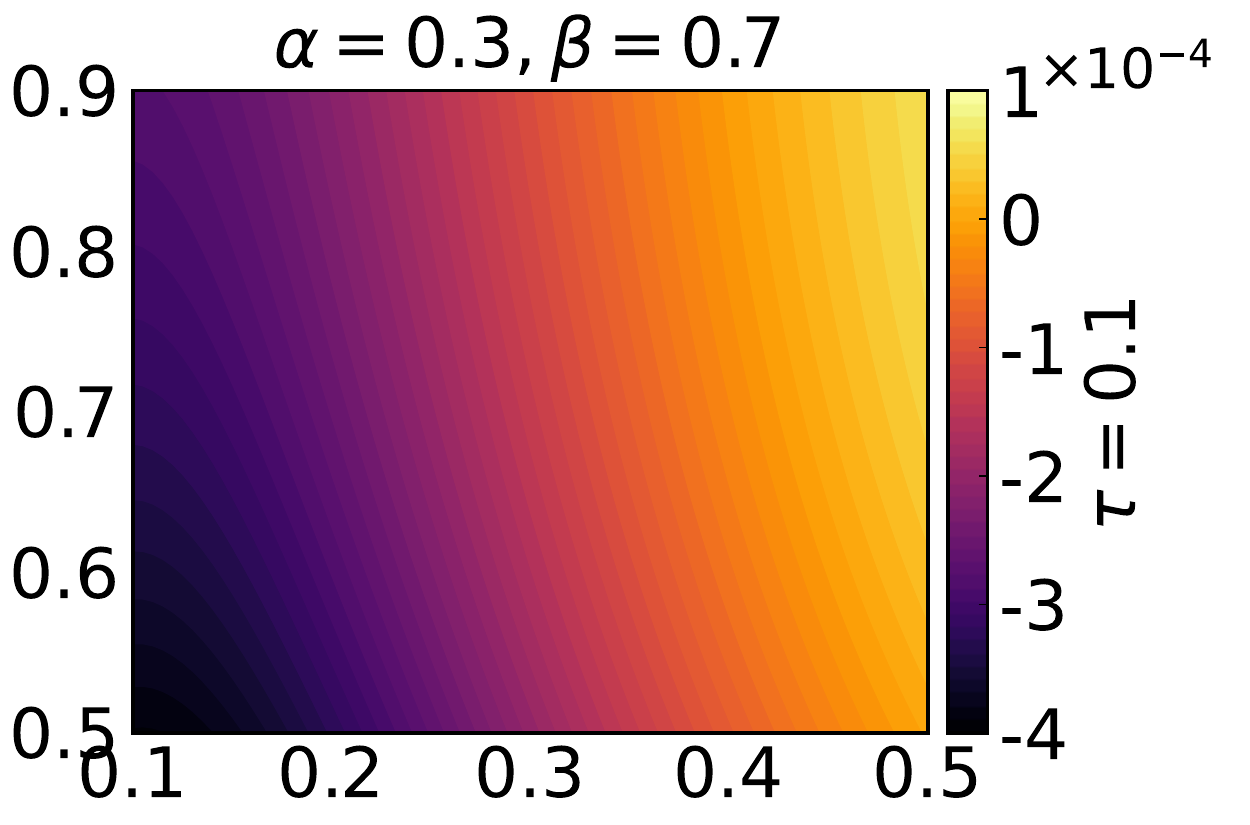}
  \vskip -.2cm
  \caption{}
  \label{fig:c3_low}
\end{subfigure}
\\
\begin{subfigure}{.3\textwidth}
  \centering
  % include first image
  \includegraphics[width=.8\linewidth]{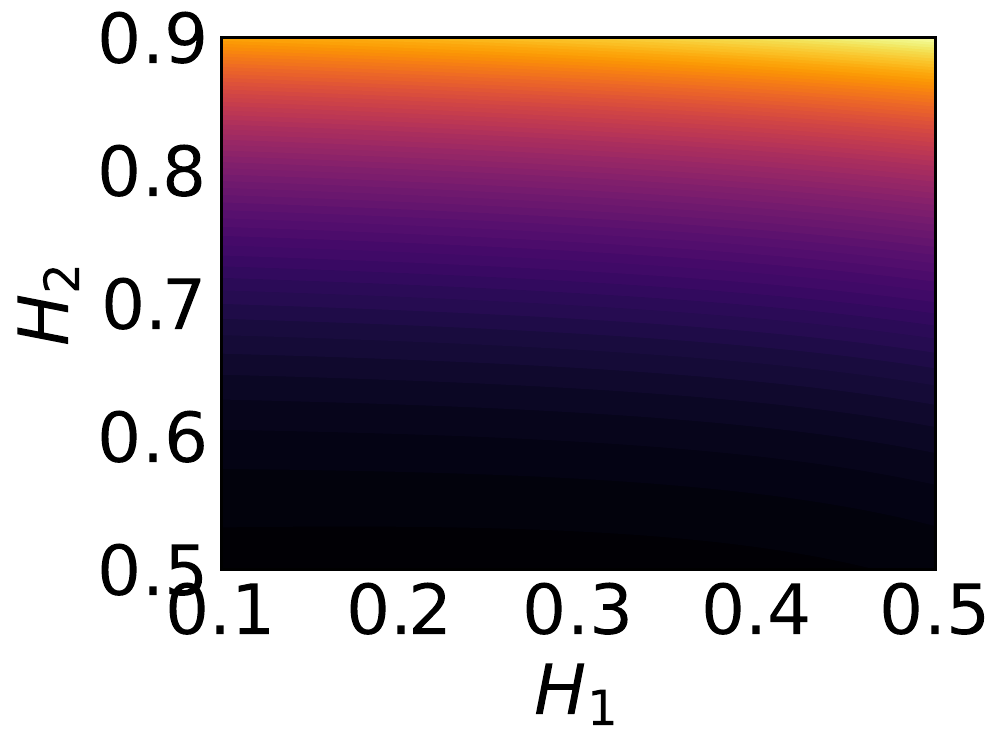}  
  \vskip -.2cm
  \caption{}
  \label{fig:c1_high}
\end{subfigure}
\begin{subfigure}{.3\textwidth}
  \centering
  % include second image
  \includegraphics[width=.8\linewidth]{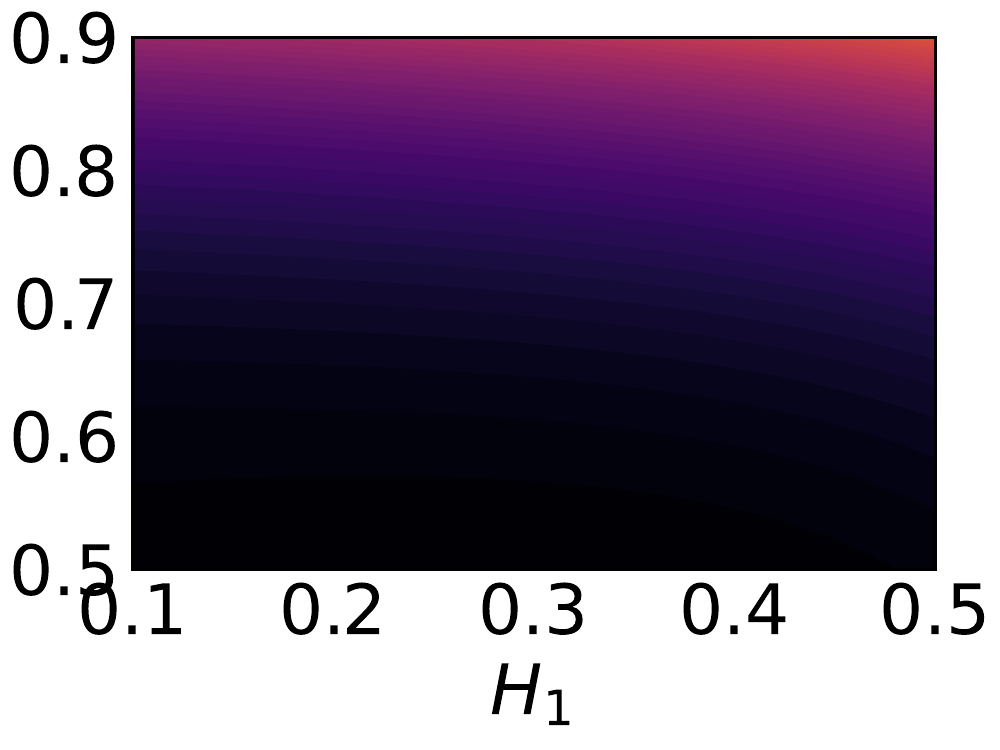}
  \vskip -.2cm
  \caption{}
  \label{fig:c2_high}
\end{subfigure}
\begin{subfigure}{.3\textwidth}
  \centering
  % include second image
  \includegraphics[width=.9\linewidth]{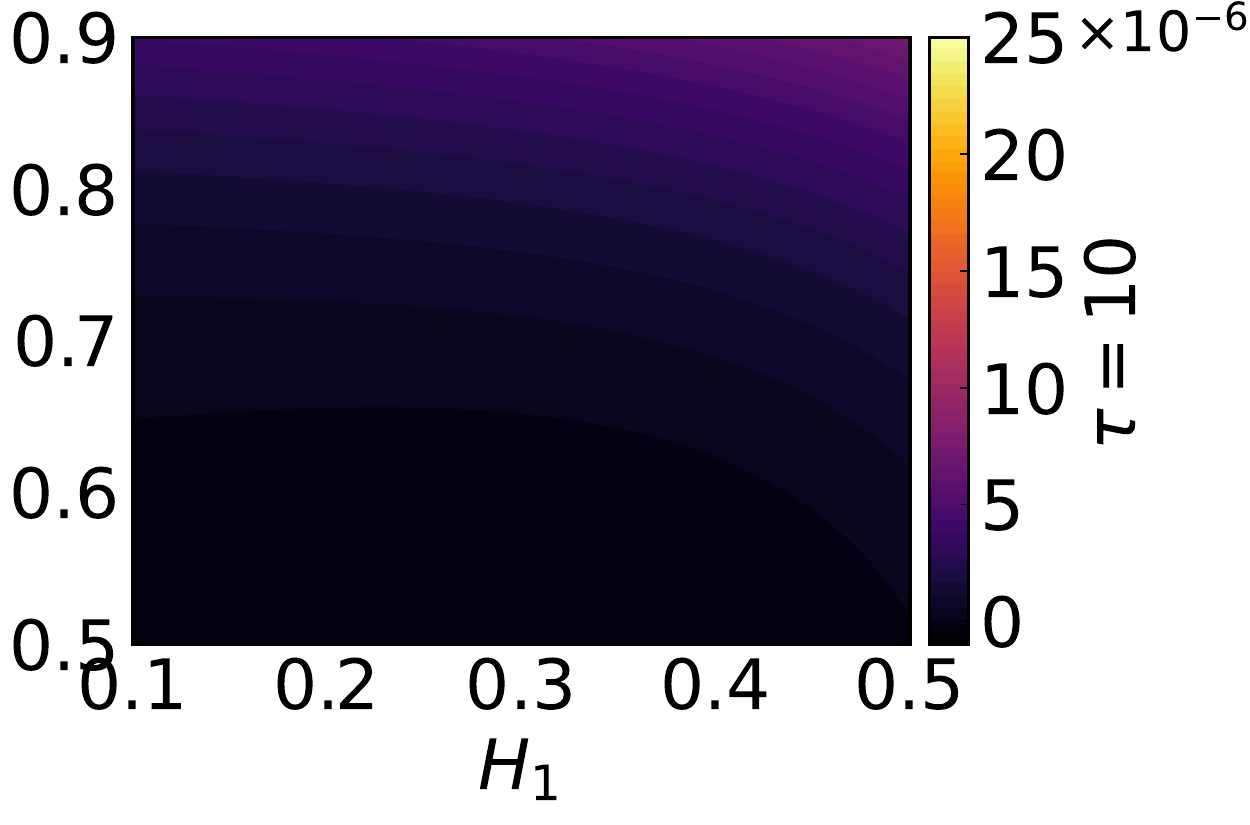}
  \vskip -.2cm
  \caption{}
  \label{fig:c3_high}
\end{subfigure}
\caption{Carpet plots of ACVFs for the FBMRE increments for $\Delta=0.01$ with the beta distribution of the Hurst exponent defined on the interval $(H_1,H_2)$, for three different cases  $\alpha=0.7$, $\beta=0.3$ (left panels), $\alpha=0.5$,  $\beta=0.5$ (middle panels), and $\alpha=0.3$, $\beta=0.7$ (right panels), and for two different lags $\tau=0.1$ (top panels) and $\tau=10$ (bottom panels). }
\label{fig:acvf_carpet_c}
\end{figure*}

To gain more insight into the persistence transition observed for the two-point and beta distributions, we analyse in detail the ACVFs $C_{\mathcal{H}}(\tau,\Delta)$ in Figures \ref{fig:acvf_carpet} and \ref{fig:acvf_carpet_c} for two different lag values $\tau=0.1$ and  $\tau=10$ which differ by two  orders of magnitude.
We can see in Figure \ref{fig:acvf_carpet}, which corresponds to the two-point distribution, that the ACVF behaviour is different for small and large lags. For  smaller lags, as probability mass $p$ at $H_1$ increases, the $H_2$ value becomes irrelevant, i.e. the carpets consist of vertical stripes. The ACVF values are predominantly positive for $p=0.1$,  and when the probability mass increases the negative values start to dominate.  In contrast, for the larger lags, the plots resemble patches of horizontal stripes especially for small probability masses at $H_1$, i.e. the $H_1$ values becomes irrelevant. We also notice that the ACVF is always non-negative in this case. Finally, we note that for $H_1=H_2=0.5$ with $p=0.5$ we recover the Brownian motion case and the ACVF is equal to zero.

In Figure \ref{fig:acvf_carpet_c} we present the behaviour of the ACVF for the beta distribution. It is similar to  that for the two-point distribution depicted in Figure \ref{fig:acvf_carpet}, except for the smaller lag when the $H_2$ parameter always has some effect on the ACVF (observed stripes are not perfectly vertical).
%Now, even for the most extreme probability masses at $H_1$ or $H_2$, there is no vertical or horizontal stripes, which means that both indexes $H_1$ and $H_2$ matter. 
%In  Figures \ref{fig:acvf_carpet}(a)-(c), so for the smaller lags, the ACVF values are always negative, whereas for $\tau=10$ (Figures \ref{fig:acvf_carpet}(d)-(f)) the values are positive, apart from the regions of large vales of $H_1$ and small values of $H_2$.

% two
% 0.25 0.75 1/2
% beta (0.25,075)
% alpha beat = 1/2
% uniform (0.25,0.75)

% fig3 acvf trzy rozne dwupuntkowe
% 0.25 0.75  A1 0.1
% 1/2 0.9
% , trzy rozne beta
% alpha beta 
% 0.7 0.3
% 0.5 0.5
% 0.3 0.7

% persistence transitionn
% przekroczenie 
% jezeli oba <0.5 to ie bedzie przekroczenia

% pionowo pierwszy dominuje
% (c)
% poziome 
% (d, e)
% h2 ma znaczenie 

% fig 5
% onne zachowanie
% beta dla malego czasu jest ciagle ujemne
% ciagle sa czarne obszary - jest ujemnie; dla wiekszych czasow bylyby juz dodatnie

%{PDFs \subref{fig:pdf-beta-long}... }}.

%{
%PDFs \subref{fig:acvf-2point} \subref{fig:acvf-beta}... }

\begin{figure*} % tau+ cov
\begin{subfigure}{.3\textwidth}
  \centering
  % include first image
  \includegraphics[width=.9\linewidth]{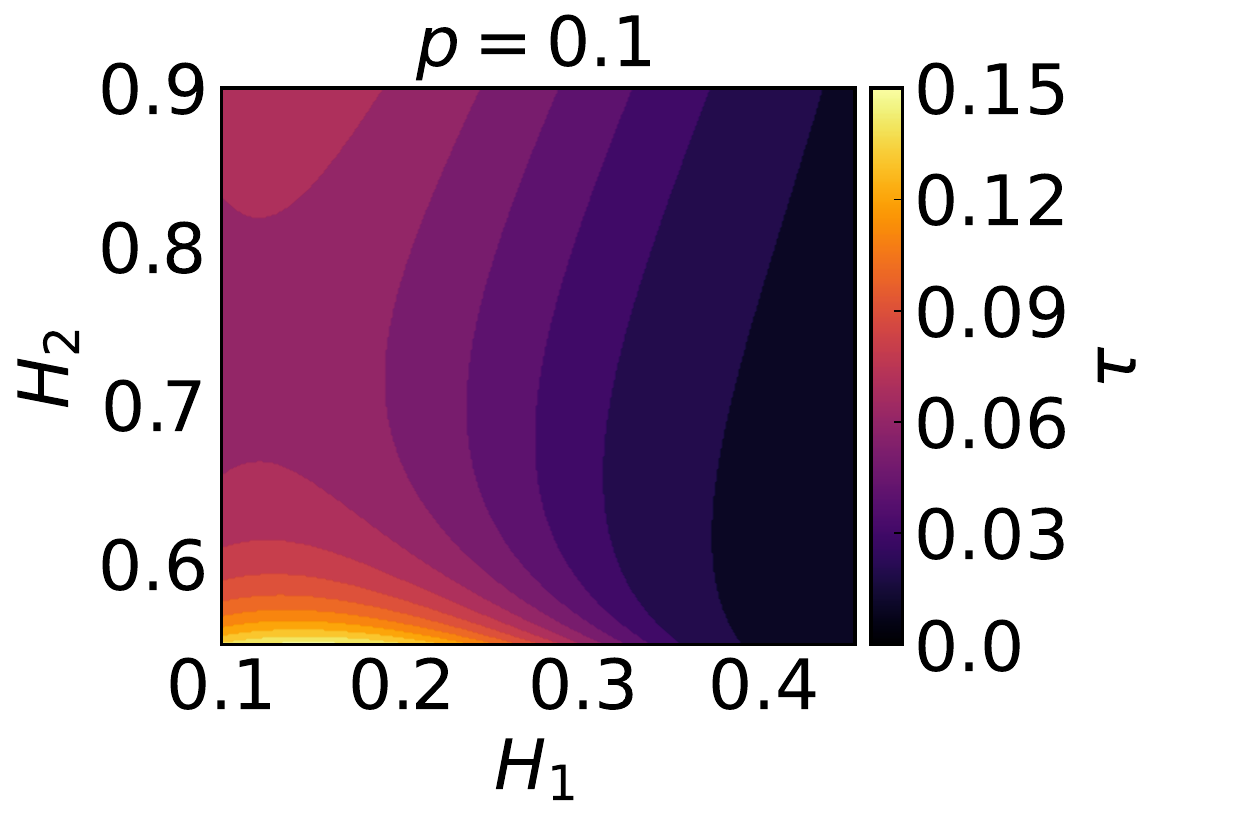}  
  \vskip -.2cm
  \caption{}
  \label{fig:tau_a1}
\end{subfigure}
\begin{subfigure}{.3\textwidth}
  \centering
  % include second image
  \includegraphics[width=.9\linewidth]{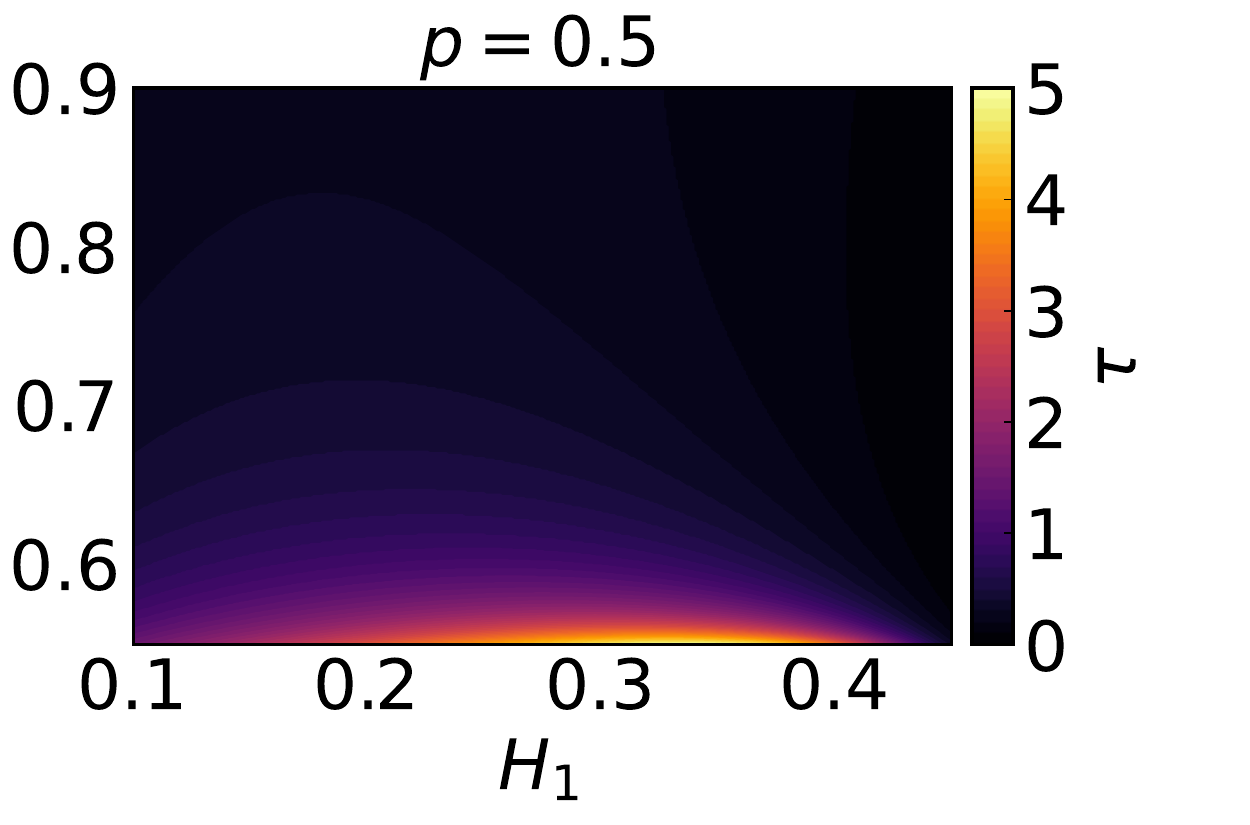}
  \vskip -.2cm
  \caption{}
  \label{fig:tau_a2}
\end{subfigure}
\begin{subfigure}{.3\textwidth}
  \centering
  % include second image
  \includegraphics[width=.9\linewidth]{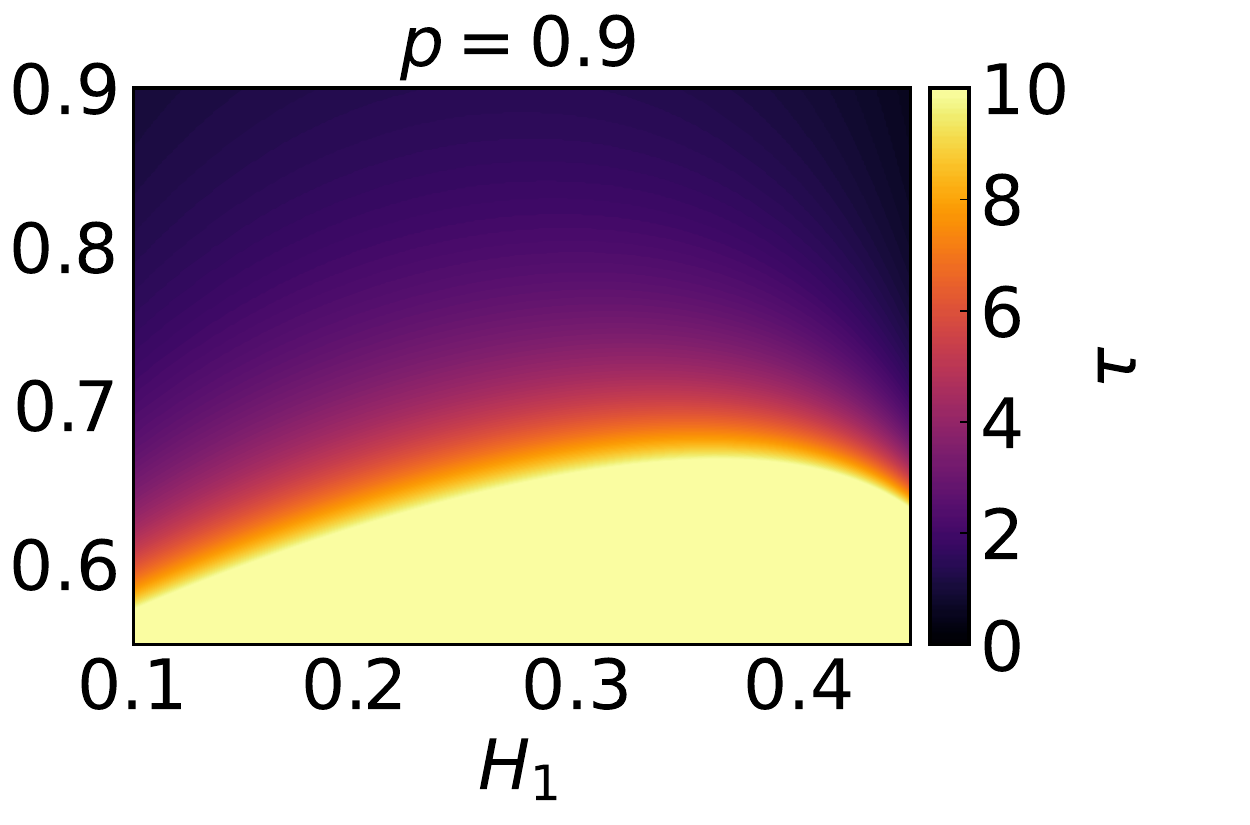}
  \vskip -.2cm
  \caption{}
  \label{fig:tau_a3}
\end{subfigure}
\caption{Carpet plots of the persistence transition lags for the two-point distribution of the Hurst index defined on the interval $(H_1,H_2)$  and $\Delta=0.01$, for three cases: $p=0.1$ (panel (a)), $0.5$ (panel (b)), and $0.9$ (panel (c)). }
\label{fig:tau_two}
\end{figure*}

\begin{figure*} % tau+ cov
\begin{subfigure}{.3\textwidth}
  \centering
  % include first image
  \includegraphics[width=.9\linewidth]{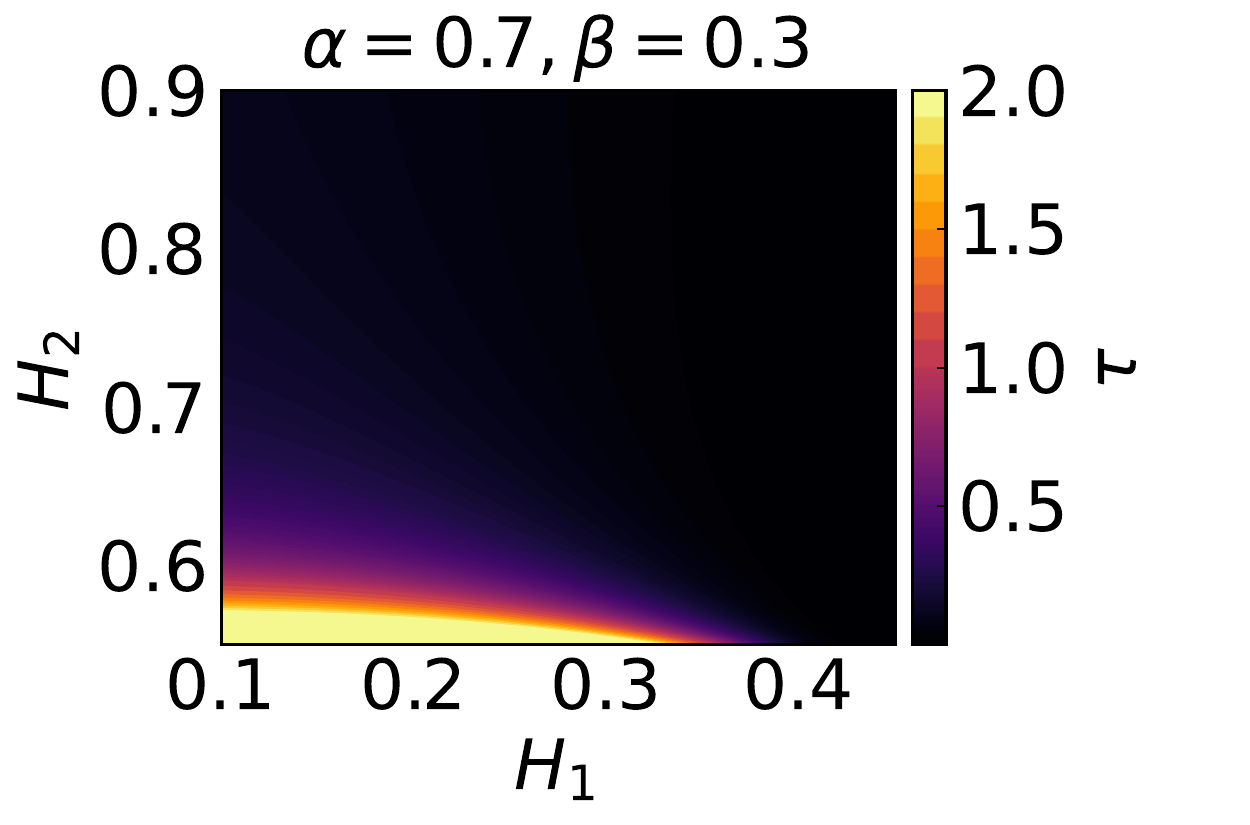}  
  \vskip -.2cm
  \caption{}
  \label{fig:tau_c1}
\end{subfigure}
\begin{subfigure}{.3\textwidth}
  \centering
  % include second image
  \includegraphics[width=.9\linewidth]{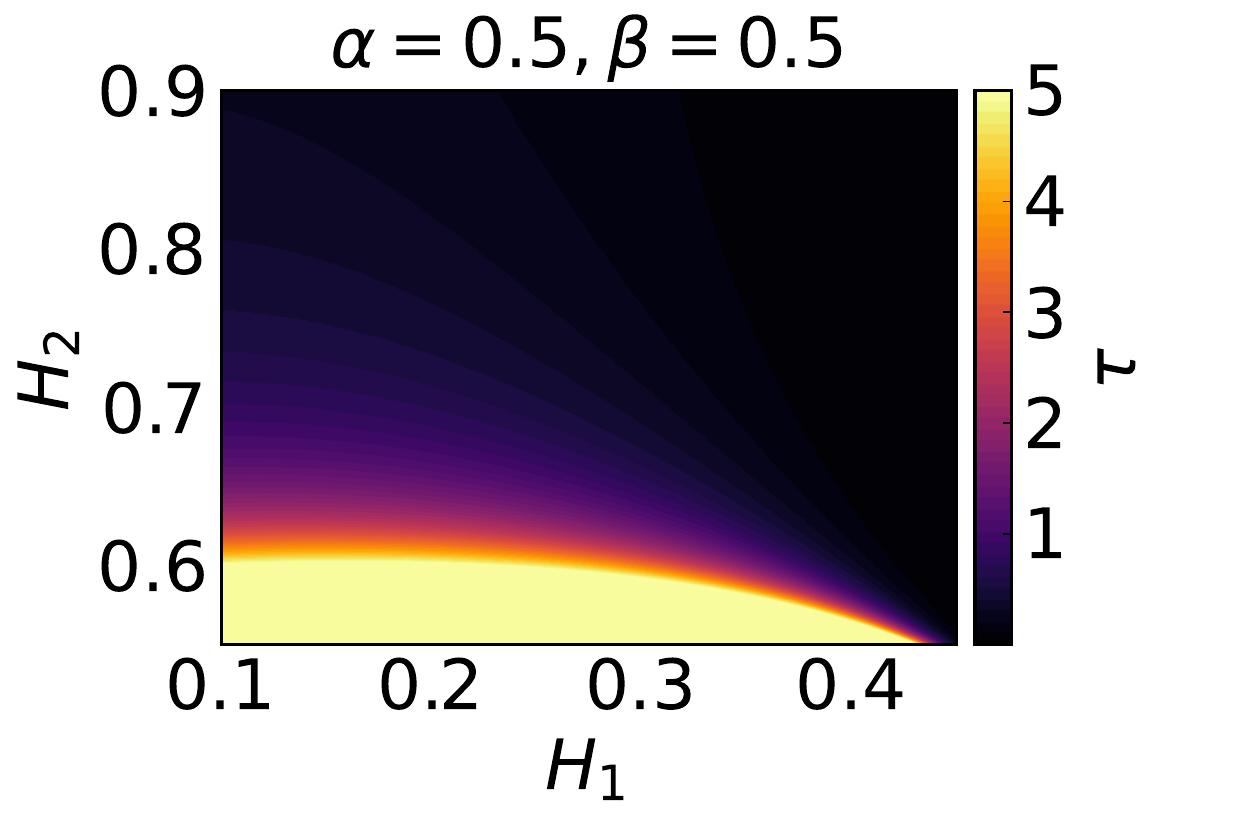}
  \vskip -.2cm
  \caption{}
  \label{fig:tau_c2}
\end{subfigure}
\begin{subfigure}{.3\textwidth}
  \centering
  % include second image
  \includegraphics[width=.9\linewidth]{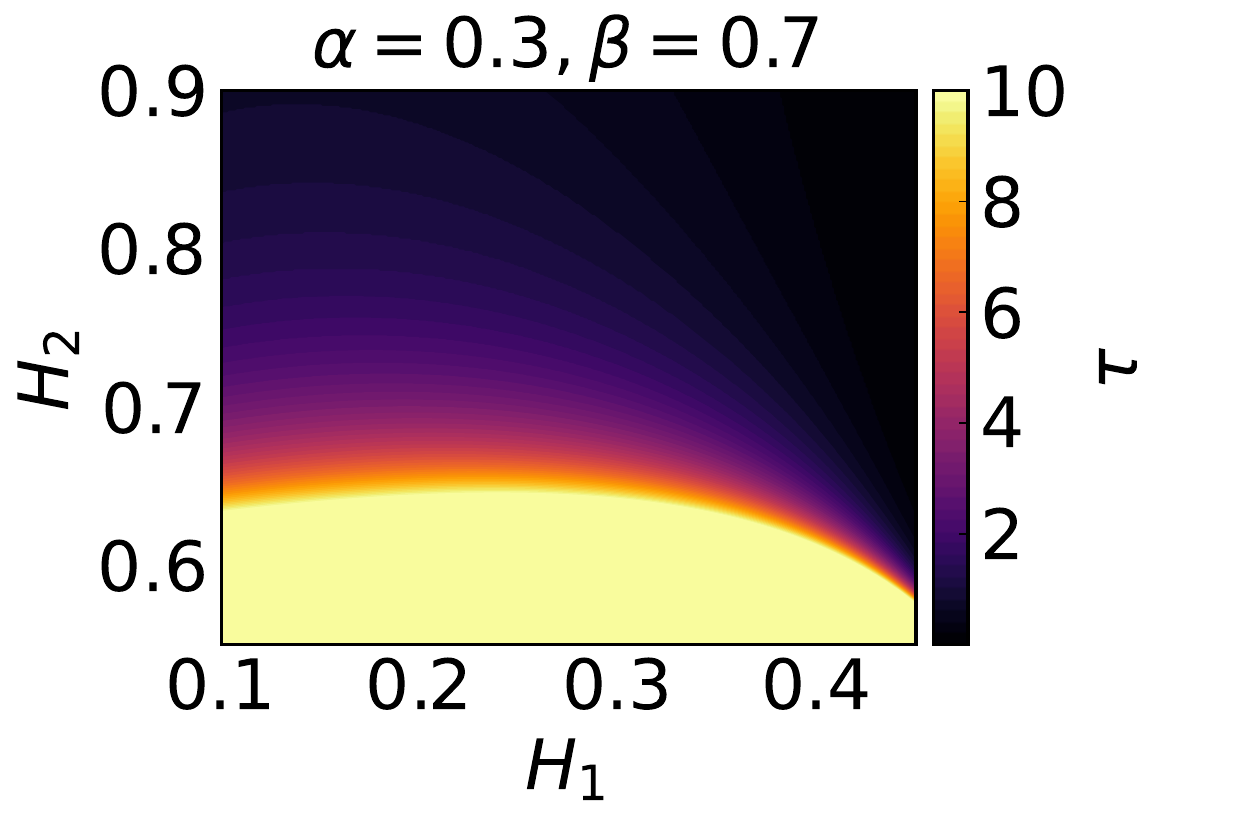}
  \vskip -.2cm
  \caption{}
  \label{fig:tau_c3}
\end{subfigure}
\caption{Carpet plots of the persistence transition lags for the beta distribution of the Hurst exponent defined on the interval $(H_1,H_2)$ and $\Delta=0.01$, for three cases: $\alpha=0.7$, $\beta=0.3$ (panel (a)), $\alpha=0.5$, $\beta=0.5$ (panel (b)), and $\alpha=0.3$, $\beta=0.7$ (panel (c)).}
\label{fig:tau_beta}
\end{figure*}

Finally, for the two considered  distributions we calculate the persistence transition lags and present them in the form of carpet plots in Figures \ref{fig:tau_two} and \ref{fig:tau_beta}.
We can observe that the waiting times for the transition are similar for the two-point and beta distributions.
%than for the two-point distribution. Moreover, for the beta distribution (Figure \ref{fig:tau_beta}) the  carpet plots look quite similar whereas for the two-point distribution (Figure \ref{fig:tau_two}) the waiting times are very different for various choices of the probability mass~$p$.
%We can conclude that the three cases of beta distribution demonstrate qualitative similarity with the corresponding three cases of two-point distribution; however quantitatively they are different. 

To sum up, numerical experiments show that the FBMRE exhibits distinct properties. First, it leads to accelerating diffusion, namely the MSD exponent  increases in time, see Figure \ref{fig:msd}. Second, the distribution of the process increments is not Gaussian: for the two-point distribution of the Hurst exponent it follows a mixture of two Gaussian laws and for the beta distribution the tails become exponential (see Figure \ref{fig:pdf}). Third, the increment process can show persistence transition, namely its ACVF switches sign at some lag. This behavior is presented in Figure \ref{fig:acvf} and is further studied in Figures \ref{fig:acvf_carpet}-\ref{fig:tau_beta}. 
%Finally, we note that although we selected parameters for the two-point and beta distributions to assure the same range and skewness type, the two distributions lead to a quantitatively very  different behaviour of the ACVF, cf. Figures \ref{fig:acvf_carpet}-\ref{fig:tau_beta}.

\section{Discussion and conclusion}
\label{sec:5}

% The FBM, introduced and Kolmogorov in 1940 \cite{kol40} and made popular by Mandelbrot and Van Ness in 1968 \cite{mannes68} is the only Gaussian self-similar process with stationary increment. For the last decades it attracted attention in many different applied fields, like, e.g. hydrology, telecommunications, economics, engineering \cite{dok-opp-taqqu} and recently in physics and biology, where it serves as a classical model for anomalous diffusion \cite{inne1,pccp}. This is mainly due to its Gaussianity and the power-law behaviour of its increment process, FGN, which for $H>1/2$ leads to the notion of long-range dependence (long memory) \cite{beran2016long}. 

% Since recent results on a number of SPT experiments in heterogeneous and crowded environments show that the Hurst exponent itself often varies from trajectory to trajectory, a natural generalisation of FBM to FBMRE is well-motivated. We note that this idea is similar to the idea of mixed Poisson process where the intensity parameter is a random variable \cite{ammeter}. Such process can be attributed to a heterogeneous portfolio of insurance policies \cite{book:grandell}. 
% The FBMRE is special case of the (MPRE), which allows the Hurst parameter to vary stochastically in time \cite{ayachetaqqu2005}. In actuarial mathematics this process corresponds to the idea of Cox (doubly- stochastic Poisson) process, where the intensity does not only depends on time but is a stochastic process \cite{coxprocess,book:grandell}. 

In this section, we discuss several issues related to the theory of FBMRE and its further extensions which we suppose to address in future studies.

\subsection{{Beta distribution fits empirical PDFs of the Hurst exponent}}  %What kind of distributions of H are observed in experimental data?}
%Fit Samu's data to beta distribution with specific parameters. Add a few words about stochastic diffusion coefficient along with random Hurst exponent.\textcolor{red}{Shall we add about random diffusion coefficient, or better hide it under the carpet ???}

\begin{figure}
    \centering
    \includegraphics[width=.9\linewidth]{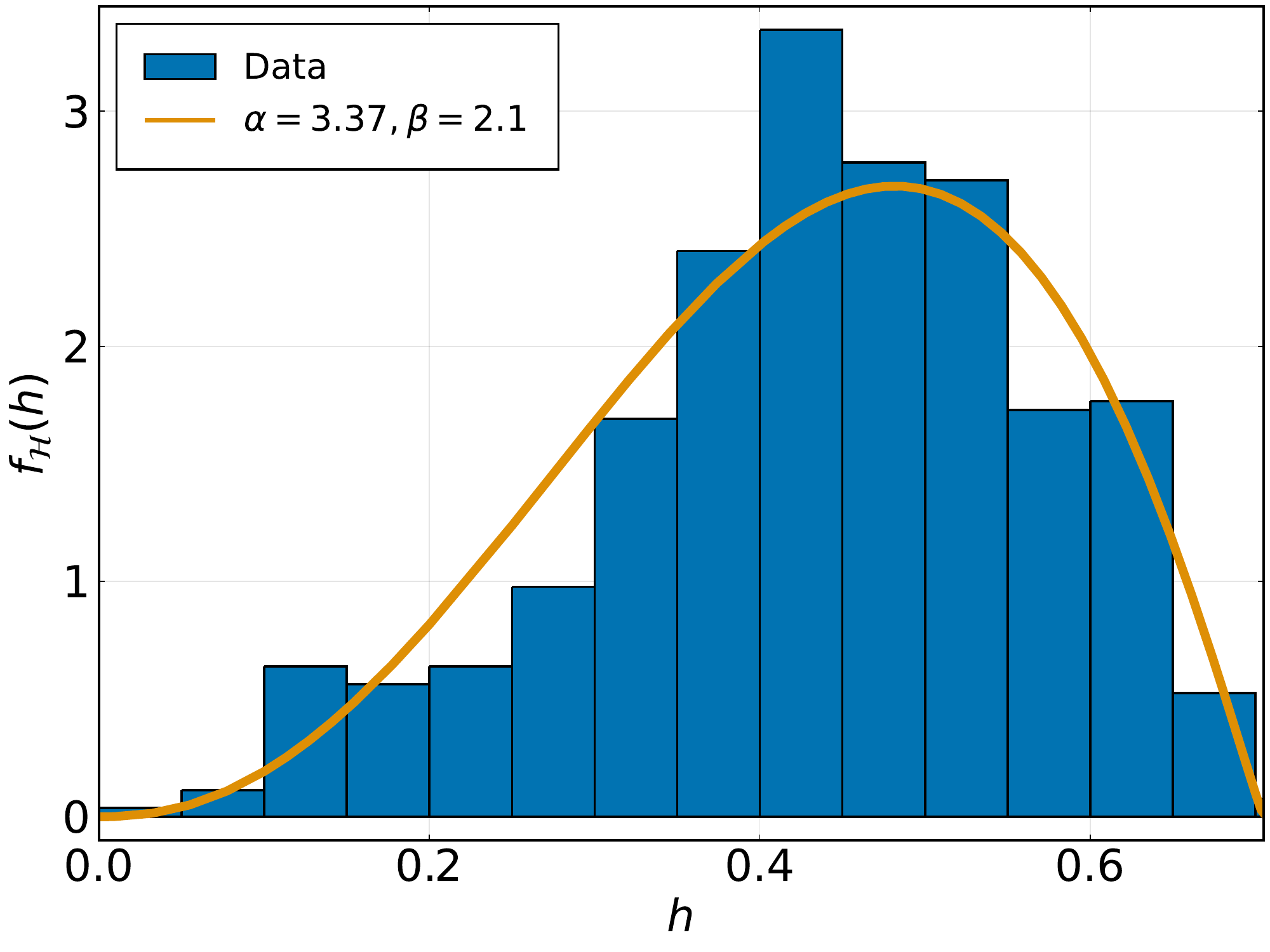}
    \caption{Histogram of Hurst exponents estimated using Bayesian inference in Ref.\cite{cherstvy2019non} for tracers tracked in mucin hydrogels and the fitted probability density function of the beta distribution. {The data set corresponds to $N=532$ two-dimensional trajectories of tracers tracked at $pH=2$ and zero salt concentration.}}
    \label{fig:exp}
\end{figure}

In single particle tracking experiments with tracers in mucin hydrogels under acidic conditions ($pH=2$), FBM was found to be the most likely model for majority  of the trajectories \cite{cherstvy2019non}. Further, Bayesian inference at the single trajectory level revealed  trajectory to trajectory fluctuations in the estimated Hurst exponent. Such variations  were found to be consistent with estimates from the TAMSD for each trajectory. In Figure \ref{fig:exp} we show that the distribution of Hurst exponents estimated using Bayesian inference in Ref.\cite{cherstvy2019non} is consistent with a beta distribution. The maximum likelihood estimators are $\alpha=3.37$ and $\beta=2.1$. We also performed the Kolmogorov-Smirnov goodness-of-fit test \cite{agostino} for the beta distribution with the estimated parameters and obtained the $p$-value equal to $0.2557$ which does not lead to rejection of the underlying beta distribution hypothesis with $5\%$ significance level. 

\begin{figure}
    \centering
    \includegraphics[width=.9\linewidth]{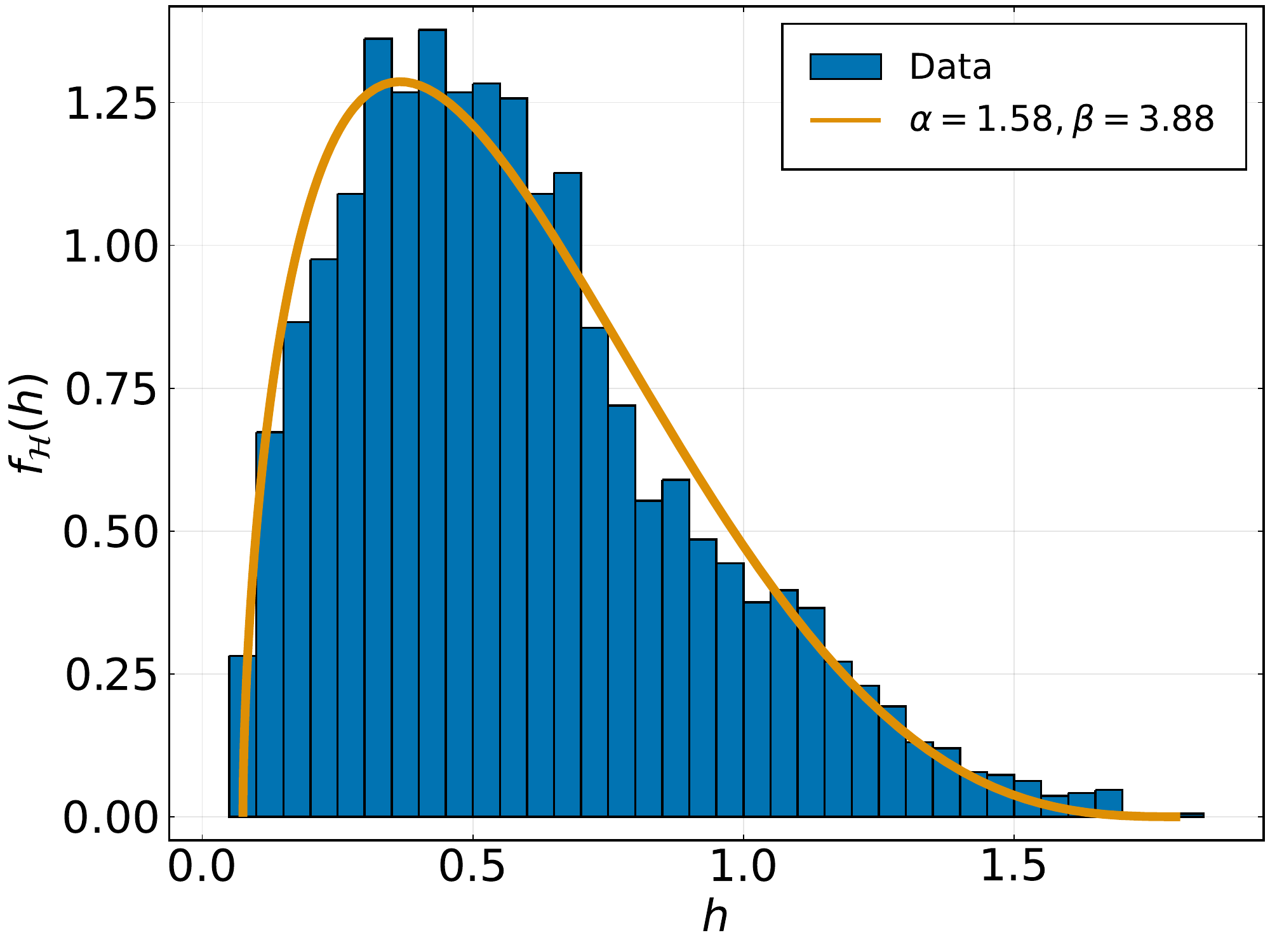}
    \caption{{Histogram of Hurst exponents estimated using MSD approach in Ref.\cite{sabri2020elucidating, janczura2021identifying} for tracer particles in the cytoplasm of mammalian cells and the fitted probability density function of the beta distribution. The data set corresponds to $N=3834$ trajectories. }}
    \label{fig:weissdata}
\end{figure}

{Another example comes from the experiment with tracer particles in mammalian cells\cite{sabri2020elucidating, janczura2021identifying}.  
The experimental observations can be described by FBM with the marked heterogeneity in individual particle trajectories. In Figure \ref{fig:weissdata} we demonstrate that the distribution of Hurst exponents is consistent with the beta distribution with $\alpha=1.58$ and $\beta=3.88$. The Kolmogorov-Smirnov goodness-of-fit test for beta distribution with the estimated parameters gives the $p$-value equal to $0.178$.}

{The two presented examples show usefulness of beta distribution in fitting and analyzing experimental data with Hurst exponent randomly changing from trajectory to trajectory. This also justifies our choice of beta distribution for the analytical studies. It would be of interest to provide similar analysis with the empirical distributions  presented in Refs.\cite{wang2018,benelli2021sub,han2020deciphering,korabel2021local}.}
%The observation in ref.\cite{cherstvy2019non} that the Hurst exponent is random from trajectory-to-trajectory with a broad distribution is in concordance with those presented in refs.\cite{wang2018,benelli2021sub,han2020deciphering,korabel2021local}. 
%This highlights, first, the need to analyze single particle tracking data using single-trajectory analysis methods and, second, to explore theoretical models incorporating the randomness of the FBM parameters.  

\subsection{Is the Hurst exponent a constant or a  random variable?}
%Krzysiek: For me what is more important is the question: Is the Hurst exponent a random variable or constant? In the article we should concentrate on this - this will the motivation for writing this paper that we observe deviations from H being constant. We do not discuss H as a stochastic process. This will be a topic for another paper.

The Hurst exponent can be estimated from experimental or simulated data using different statistics such as  e.g., R/S statistic, rescaled variance statistic, detrended fluctuation analysis or TAMSD \cite{CAJUEIRO2005172,MIELNICZUK20074510,Carbone_2007,ELLIS2007159,Sikora_2017,Sikora_2017_2}. As a result, one obtains the value of the estimator $\hat{H}$ that is a random variable. The papers \cite{ELLIS2007159,CAJUEIRO2005172} discuss the sample distribution of $\hat{H}$ while the analytical derivation of the distribution of $\hat{H}$ for a finite number of data points was provided in \cite{Sikora_2017} by using TAMSD statistic for FBM.  How can one  distinguish between two different sources of randomness of the estimated Hurst exponent, namely  finiteness of the data set and random change from trajectory to trajectory? This issue points to the need to introduce effective tools for distinguishing between these two cases. One  idea could be to measure the MSD as a function of time which should demonstrate either linear  or convex behavior (in log-log scale)  in time  for a single  or random Hurst exponent, respectively. Another proposition is to compare the sample distribution obtained for the estimated Hurst exponent by using TAMSD statistic with the analytical distribution presented in \cite{Sikora_2017}. Alternatively, the uncertainty in the estimated value of the parameter due to finiteness of the data  can be quantified using single trajectory analysis methods such as Bayesian inference and machine learning. Bayesian inference provides with the posterior distribution of the parameters---for example the Hurst exponent---given the data and a prior on the parameters. The variance of such a posterior distribution then quantifies the uncertainty in the estimated value of the parameter. Performing Bayesian parameter estimation on simulated trajectories of different lengths can then quantify the uncertainty in the estimation due to the finite length of trajectories. Such studies have been conducted in, for example, Refs.\cite{thapa2018,park2021,thapa2022}. The quantification of uncertainty due to finite number of data points has also been done using machine learning approaches \cite{andi21}. 

%paragraph by Agnes and Aleksei:
% The Hurst exponent can be estimated from experimental or simulated data using different statistics such as  e.g., R/S statistic, rescaled variance statistic, detrended fluctuation analysis or time average mean square displacement (TAMSD) \cite{CAJUEIRO2005172,MIELNICZUK20074510,Carbone_2007,ELLIS2007159,Sikora_2017,Sikora_2017_2}. As the result one obtains the value of the estimator $\hat{H}$ that is a random variable. The papers \cite{ELLIS2007159,CAJUEIRO2005172} discuss the sample distribution of $\hat{H}$ while the analytical derivation of the distribution of $\hat{H}$ for the finite number of data points was provided in \cite{Sikora_2017} by using TAMSD statistic for FBM.  So, how to distinguish between two different sources of randomness of the estimated Hurst exponent, namely  finiteness of the data set and heterogeneity of the medium? To the best of the authors' knowledge this issue was not addressed in the literature so far. Thus, there is a need to introduce an effective tools for distinguishing between such two cases. One of the idea could be to measure the MSD as a function of time which should demonstrate either linear  or convex behavior in time for a single  or random Hurst exponent, respectively. Another proposition is to compare the sample distribution obtained for the estimated Hurst exponent by using TAMSD statistic with the analytical distribution presented in \cite{Sikora_2017}. 

\subsection{Is the Hurst exponent a random variable or a  random process?}

Recent studies of the random intracellular motions point to the necessity of going  beyond the FBMRE concept  to explain the highly heterogeneous transport of structures inside living cells \cite{han2020deciphering,korabel2021local,Korabel_arxiv}. The MPRE allowing the Hurst parameter to vary
stochastically in time \cite{ayachetaqqu2005} is a natural extension of FBMRE, and its basic mathematical properties are worth to be explored, similar to how it is done in  the present paper. We also note that the two processes, FBMRE and MPRE, can be distinguished by analyzing their TAMSD, namely while FBMRE is a mean-squared ergodic process, the MPRE does not exhibit this property.

\subsection{Mimicking heterogeneous environment with random Hurst exponent}
The anomalous transport in biological media has largely been attributed to heterogeneity \cite{franosch2013} which itself can arise from two sources: (a) the spatially heterogeneous macromolecular crowding in, for instance, the cell cytoplasm and/or (b) heterogeneity in the shape and size of different constituents, such as proteins and lipids, of a biological cell. The former source may correspond to having local patches in a viscoelastic environment where a synthetic particle of known shape and size---tracked for a finite time such that it diffuses within the patch during this period---diffuses with a constant Hurst exponent. Another particle in a neighbouring patch in the medium might diffuse with a different Hurst exponent. Thus, the random Hurst exponents extracted from the trajectories of such particles mimic the heterogeneous environment.  However, when the tracked particles are cellular constituents, such as proteins, the ensemble itself might be heterogeneous (non-identical particles), potentially giving rise to trajectory to trajectory fluctuations in the estimates of the Hurst exponent. Besides, such fluctuations may also arise from the superimposition of active processes on the motion of tracked particles \cite{benelli2021sub}. How to disentangle the two potential sources of heterogeneity---listed above as (a) and (b)---is an interesting research question which might be investigated experimentally. 
We note here that the authors in Ref.\cite{manzo2015} explained the dynamics of a transmembrane protein (DC-SIGN) on living-cell membranes by considering local patches of fixed diffusivity, which mimicked the heterogeneity of the cell membranes. 

FBMRE presented in this article describes the scenarios of heterogeneity discussed in the preceding paragraph.  
However, the scenarios can be further complicated if, for example, the local patches with fixed Hurst exponent are small such that a tracked particle transitions multiple times between such patches with different Hurst exponents during the total experimental time period. Moreover, when the tracked particle is a cell-constituent such as a protein, it may change shape and size during its dynamics. Furthermore, given that a biological cell is dynamic, the cellular environment itself might change during the experiment. These scenarios would call for a generalization of FBMRE to include a Hurst exponent which is stochastic along an individual trajectory. We are currently exploring such generalizations as a follow-up of the present article.

%Moreover, the cellular environment itself is dynamic, evolving in time.  

In summary, in this paper we study the basic mathematical properties of the fractional Brownian motion with the Hurst exponent randomly changing from trajectory to trajectory. {To the best of our knowledge, this is the first paper in the literature providing such complex analysis of the process with random Hurst exponent.} Specifically, we provide general expressions for the probability density function, the $q-$th moment (including  mean squared displacement) and  autocovariance function for the increment process. We derive explicit results for three generic distributions of random Hurst exponent, namely,  two-point, uniform and beta distributions. Our analytical and numerical analyses reveal two effects which are the hallmarks of the process, namely accelerating diffusion and persistence transition. {Taking into account that three generic types of the distributions considered in our paper exhibit qualitatively similar properties, we conclude that these two effects are quite common for FBMRE. Our analytical expressions quantify and illustrate how they arise, and our simulations validate them.} The presented results pave the way to a consistent mathematical and statistical description of fractional motions in complex heterogeneous environment. {We hope that this paper will be of interest to researchers working not only in the field of biology but also climate, hydrology and financial markets where FBM is one of the canonical processes. Moreover, we hope to inspire experimentalists to test the two phenomena we discovered. Given the number of recent experiments which have reported FBMRE, we believe that the mathematical framework and the numerical methods we present are of significant potential utility.}
%\section{Conclusions}
%Summary of the paper, short paragraph.
\begin{acknowledgments}
KB would like to acknowledge the Beethoven Grant No. DFG-NCN 2016/23/G/ST1/04083. ST acknowledges support in the form of a Sackler postdoctoral fellowship and funding from the Pikovsky-Valazzi matching scholarship, Tel Aviv University.
The work of AW was supported by National Center of Science under Opus Grant 2020/37/B/HS4/00120 ``Market risk model identification and validation using novel statistical, probabilistic, and machine learning tools''. AC acknowledges the support of the Polish National Agency for Academic Exchange (NAWA). 

\end{acknowledgments}

\section*{Data Availability}
The data that support the findings of this study are available from the corresponding author upon reasonable request.
\appendix
\section{Derivation of $A_H$ prefactor}\label{appA}
Starting with  the integral representation of FBM as given in Eq.~(\ref{FBM_main}), we get

\begin{eqnarray}
\label{eq-msd-a1}
\mathbb{E}\left(B_{H}^2(t)\right) &=&A^2_{H}t^{2H}\Bigg\{\frac{1}{2H}\nonumber\\
 &+&\int_{0}^{\infty}\left[(1+u)^{H-1/2}-u^{H-1/2}\right]^2 du \Bigg\}.\quad
\end{eqnarray}
The formula in parentheses
can be written as
\begin{eqnarray}
\label{eq-dera1}
\Bigg\{ \ldots \Bigg\}&=& \frac{1}{2H} + \lim_{a \to \infty}\int_{0}^{a} (1+u)^{2H-1}du\nonumber\\
 &+& \lim_{a \to \infty}\int_{0}^{a} \; u^{2H-1}du \nonumber\\
 &-&2\lim_{a \to \infty}\int_{0}^{a} \; u^{2H-1}\left( 1+\frac{1}{u}\right)^{H-1/2}du \nonumber \\
 &=& \lim_{a \to \infty}\left[\frac{a^{2H}}{H}+a^{2H-1}-2I(a,H) \right],
\end{eqnarray}
where
\begin{equation}
\label{eq-dera2}
   I(a,H)= \int_{0}^{a} \; u^{2H-1}\left(1+\frac{1}{u}\right)^{H-1/2}du
\end{equation}
is the incomplete beta function.

Now, we take the integral $I(a,H)$ twice by parts  in order to extract the terms that diverge at $a\to \infty$,
\begin{eqnarray}
\label{eq-dera3}
   I(a,H)&=& \frac{a^{2H}}{2H}+\frac{a^{2H-1}}{2} \nonumber \\
   &+&\frac{H-3/2}{4H}\int_0^a \; u^{2H-3}\left(1+\frac{1}{u}\right)^{H-5/2} du. \quad
\end{eqnarray}
Note that the integral in Eq.~(\ref{eq-dera3}) converges at $a\to \infty$. After plugging Eq.~(\ref{eq-dera3})
into Eq.~(\ref{eq-dera1}) we obtain
\begin{eqnarray}
\label{eq-dera4}
\Bigg\{ \ldots \Bigg\}&=& \frac{3/2-H}{2H} \int_0^{\infty} \frac{u^{H-1/2}}{(1+u)^{5/2-H}}du\nonumber\\
 &=& \frac{3/2-H}{2H} \mathbb{B}\left(H+\frac{1}{2},2-2H\right), \quad
\end{eqnarray}
where $\mathbb{B}(x,y)$ is the beta function defined in Eq.~(\ref{beta_fun}).
Finally, we plug Eq.~(\ref{eq-dera4}) into Eq.~(\ref{eq-msd-a1}) to obtain
\begin{equation}
\label{eq-msd-a2}
\mathbb{E}\left(B_{H}^2(t)\right) =A^2_{H}t^{2H}\frac{\Gamma^2(H+1/2)}{\Gamma(1+2H)\sin\left(\pi H\right)}.
\end{equation}
Imposing $\mathbb{E}\left(B_{H}^2(t)\right) =t^{2H}$ gives the expression for $A_H$,
\begin{equation}
    A_H=\frac{\sqrt{\Gamma(1+2H)\sin\left(\pi H\right)}}{\Gamma(H+1/2)}.
\end{equation}
\section{Asymptotics of ACVF in case of uniform distribution of the random Hurst exponent}\label{app_unif}
For small  $\tau/\Delta\ll 1$ we have the following asymptotic behavior for ACVF $C_{{\mathcal{H}}}(\tau,\Delta)$
  \begin{eqnarray}C_{{\mathcal{H}}}(\tau,\Delta)  & \sim & \frac{1}{(H_2-H_1)\log(\Delta)}\Bigg[\Delta^{2H_2}\left(1-\left(\frac{\tau}{\Delta}\right)^{2H_2}\right)\nonumber\\
  &-&\Delta^{2H_1}\left(1-\left(\frac{\tau}{\Delta}\right)^{2H_1}\right)\Bigg].
       \end{eqnarray}   
 Note that  $C_{\mathcal{H}}(0,\Delta)$ is always positive, as it should be. For large  $\tau/\Delta\gg 1$ the Eq.~(\ref{acvf_unif}) gets a simpler form if in addition we require one of the following conditions, namely either a) $\Delta\ll \tau\ll 1$ or b) $\tau\gg \Delta\gg 1$. Then we get 
 \begin{eqnarray}C_{{\mathcal{H}}}(\tau,\Delta)  & \sim & \frac{\Delta^2}{(H_2-H_1)\log(\tau)} \left(H_2(2H_2-1)\tau^{2(H_2-1)} \right. \nonumber \\
    &-& \left. H_1(2H_1-1)\tau^{2(H_1-1)}\right)+\delta C_{\mathcal{H}},
    \end{eqnarray} 
and the correction term $\delta C_{\mathcal{H}}$ contains additional prefactor $1/\log(\tau)$. \section{Beta distribution on the interval $(0,1)$ of the random Hurst exponent}\label{app1}
The PDF of beta distribution on the interval $(0,1)$ with parameters $\alpha>0$ and $\beta>0$ reads \begin{eqnarray}\label{beta_pdf}f_{\mathcal{H}}(h)=\frac{h^{\alpha-1}(1-h)^{\beta-1}}{\mathbb{B}(\alpha,\beta)}{I}_{h\in (0,1)}.\end{eqnarray}
The MFG is given by
   \begin{eqnarray}\label{beta_mgf}
      M_{\mathcal{H}}(s) = {}_1 F_1(\alpha, \alpha+\beta, s),
   \end{eqnarray}
   where $\mathbb{B}(\alpha,\beta)$ is the beta function defined in Eq.~(\ref{beta_fun}), and ${}_1 F_1(\cdot, \cdot, \cdot)$ is a confluent hypergeometric function (Kummer function)
        \begin{eqnarray}\label{F1function}
            {}_1F_1(a, b, z) = \sum_{n=0}^\infty \frac{(a)_n}{(b)_n} \frac{z^n}{n!} = 1+\frac{a}{b}z+\cdots.
        \end{eqnarray}
    In the formula above $a, b$ are positive constants,  $z \in \mathbb{R}$, and $(a)_n$ is the Pochhammer symbol, i.e. $(a)_0 = 1$ and $(a)_n = a(a+1)(a+2)\ldots(a+n-1)$ for $n\geq 1$. 

Using Eq. (\ref{for1}) we obtain the PDF  $f_{B_{\mathcal{H}}}(x,t)$,
\begin{eqnarray}
 f_{B_{\mathcal{H}}}(x,t) &=&\frac{1}{\mathbb{B}(\alpha,\beta)}\\&\times&\int_0^1 \frac{h^{\alpha-1}(1-h)^{\beta-1}}{\sqrt{2\pi t^{2h}}} \exp\left\{\frac{-x^2}{2t^{2h}} \right\}dh.\nonumber
\end{eqnarray}

With the use of Eq. (\ref{for2}) the MSD takes the form 
    \begin{eqnarray}\label{app1_msd} %\mathbb{E}\left[B_{\mathcal{H}}^q(t)\right]&=&c_q{}_1 F_1(\alpha, \alpha+\beta, q\log( t)),\\
    \mathbb{E}\left(B_{\mathcal{H}}^2(t)\right)&=&{}_1 F_1(\alpha, \alpha+\beta, 2\log( t)).
\end{eqnarray}
Notice that for large $z$, except when $a=0, -1,\ldots$ (polynomial cases), the confluent hypergeometric function given in (\ref{F1function}) behaves as 
\begin{eqnarray}\label{sim1}
 {}_1F_1(a, b, z) \sim \frac{\Gamma(b)e^z  z^{a-b}}{\Gamma(a)}.
\end{eqnarray}
Thus, using the Kummer's transformation for the confluent  hypergeometric  function
\begin{eqnarray}\label{special_function_small}
{}_1F_1(a,b,z)=e^{z}{}_1F_1(b-a,b,-z),
\end{eqnarray}
we obtain the asymptotics of MSD for short times
\begin{eqnarray}\label{beta_small_1}
 \mathbb{E}\left(B_{\mathcal{H}}^2(t)\right)&\sim& \frac{\Gamma(\alpha+\beta)}{\Gamma(\beta)}\frac{1}{(2\log(1/t))^{\alpha}},
\end{eqnarray}
whereas for long times we have
 \begin{eqnarray} \label{beta_large_1}
    \mathbb{E}\left(B_{\mathcal{H}}^2(t)\right)&\sim&\frac{\Gamma(\alpha+\beta)}{\Gamma(\alpha)}\frac{t^2}{(2\log(t))^{\beta}}.
\end{eqnarray}
Using Eq. (\ref{for4}) and the MGF  (\ref{beta_mgf}) we obtain the ACVF for $\{b_{\mathcal{H}}^{\Delta}(t)\}$,
\begin{eqnarray}\label{acvf_beta_zero_one}
C_{{\mathcal{H}}}(\tau,\Delta) &=&\frac{1}{2} {}_1 F_1(\alpha, \alpha+\beta,2\log(\tau+\Delta))\nonumber\\&+&\frac{1}{2}{}_1 F_1(\alpha, \alpha+\beta,2\log( |\tau-\Delta|))\nonumber\\
&-& {}_1 F_1(\alpha, \alpha+\beta,2\log(\tau)).\end{eqnarray}
%\textcolor{red}{This need to be carefully checked and improved
%Thus, by using (\ref{sim1}) we have the following asymptotic behavior when $\tau\gg\Delta$ 
%\begin{eqnarray}
%C_{{\mathcal{H}}}(\tau,\Delta) &\sim&\frac{\Gamma(\alpha+\beta)}{2\Gamma(\alpha)}\left[\frac{(\tau+\Delta)^2}{(2\log(\tau+\Delta))^{\beta}}+ \frac{(\tau-\Delta)^2}{(2\log(\tau-\Delta))^{\beta}}\right]\nonumber\\
%&-&\frac{\Gamma(\alpha+\beta)\tau^2}{\Gamma(\alpha)(2\log(\tau))^\beta}.\end{eqnarray}}

\section{Asymptotics of ACVF in case of beta distribution on the interval $[H_1,H_2]$ of the random Hurst exponent}\label{app22}
For asymptotic behavior of ACVF $C_{\mathcal{H}}(\tau,\Delta)$, Eq. (\ref{acvf_beta_h1h2}), at $\tau/\Delta\ll 1$ we get
%\begin{widetext}
\begin{eqnarray}
C_{\mathcal{H}}(\tau,\Delta)&\sim&\Delta^{2H_1}\Bigg[{}_1F_1\left(\alpha,\alpha+\beta,2(H_2-H_1)\log \Delta\right)\\&-&\left(\frac{\tau}{\Delta}\right)^{2H_1}{}_1F_1\left(\alpha,\alpha+\beta,2(H_2-H_1)\log\tau\right)\Bigg].\nonumber
\end{eqnarray}
%\end{widetext}
This result acquires an elegant form in two limit cases, namely a) $\tau\ll\Delta\ll 1$, where
\begin{eqnarray}
C_{\mathcal{H}}(\tau,\Delta)&\sim&\frac{\Gamma(\alpha+\beta)}{\Gamma(\beta)[2(H_2-H_1)]^{\alpha}}\frac{\Delta^{2H_1}}{\log^{\alpha}(1/\Delta)}\nonumber\\
&\times&\Bigg[1 -\left(\frac{\tau}{\Delta}\right)^{2H_1}\frac{\log^{\alpha}(1/\Delta)}{\log^{\alpha}(1/\tau)}\Bigg]
\end{eqnarray}
and b) $\Delta\gg \tau\gg 1$ where
\begin{eqnarray}
C_{\mathcal{H}}(\tau,\Delta)&\sim&\frac{\Gamma(\alpha+\beta)}{\Gamma(\beta)[2(H_2-H_1)]^{\beta}}\frac{\Delta^{2H_2}}{\log^{\beta}(\Delta)}\nonumber\\
&\times&\Bigg[1 -\left(\frac{\tau}{\Delta}\right)^{2H_2}\frac{\log^{\beta}(\Delta)}{\log^{\beta}(\tau)}\Bigg].
\end{eqnarray}
At $\tau/\Delta\gg 1$ the asymptotics of ACVF reads 
\\\\
\begin{widetext}
\begin{eqnarray}\label{beta_app3}
C_{\mathcal{H}}(\tau,\Delta)&\sim&\frac{\Delta^{2}}{\tau^{2-2H_1}}\Bigg[H_1(2H_1-1){}_1 F_1\left(\alpha, \alpha+\beta,2(H_2-H_1)\log\tau\right)\nonumber\\&+&\frac{(H_2-H_1)(2H_1-1)\alpha}{\alpha+\beta}{}_1 F_1\left(\alpha+1, \alpha+\beta+1,2(H_2-H_1)\log\tau\right)\nonumber\\&+&\frac{(H_2-H_1)^2\alpha(\alpha+1)}{(\alpha+\beta)(\alpha+\beta+1)}{}_1 F_1\left(\alpha+2, \alpha+\beta+2,2(H_2-H_1)\log\tau\right)\Bigg].
\end{eqnarray}
\end{widetext}
%This formula acquires a much simpler form if, in addition, we consider a) $\Delta\ll \tau\ll 1$, where
%\textcolor{red}{here formula}\\
%and b) $\tau\gg\Delta\gg 1$, where \textcolor{red}{here formula}.

%% joint carpets
% \begin{figure*}[ht] 
%   \centering
%   \includegraphics[width=.9\linewidth]{carpet cov unif dt=0.01.pdf}  
%   \caption{Carpet uniform $C_\mathcal{H}(t, \Delta=0.01)$.}
%   \label{fig:carpet-cov-unif}
  
%   \includegraphics[width=.9\linewidth]{carpet cov beta dt=0.01.pdf}
%   \caption{Carpet beta $C_\mathcal{H}(t, \Delta=0.01)$.}
%   \label{fig:carpet-cov-beta}
% \end{figure*}

%\section{Second proof for Beta distribution  ACVF}\label{app3}\begin{widetext}
%\begin{eqnarray}\label{beta_app3}
%C_{\mathcal{H}}(\tau,\Delta)&\sim&\frac{\Delta^{2}}{\tau^{2-2H_1}}\Bigg[H_1(2H_1-1){}_1 F_1\left(\alpha, \alpha+\beta,2(H_2-H_1)\log\tau\right)\nonumber\\
%&+&\frac{(H_2-H_1)(2H_1-1)\alpha}{\alpha+\beta}{}_1 F_1\left(\alpha+1, \alpha+\beta+1,2(H_2-H_1)\log\tau\right)\nonumber\\
%&+&\frac{(H_2-H_1)^2\alpha(\alpha+1)}{(\alpha+\beta)(\alpha+\beta+1)}{}_1 F_1\left(\alpha+2, \alpha+\beta+2,2(H_2-H_1)\log\tau\right)\Bigg].
%\end{eqnarray}
%\end{widetext}
%\bibliographystyle{plain}
\bibliography{bibliography}

\end{document}